\documentclass[12pt, preprint]{aastex}

\usepackage{emulateapj5}

\newcommand{\grsmq}{GRS~1915$+$105}
\newcommand{\gromq}{GRO~J1655$-$40}
\newcommand{\gxbh}{GX~339$-$4}
\newcommand{\cygxone}{Cyg~X$-$1}
\newcommand{\cygxthree}{Cyg~X$-$3}
\newcommand{\rxte}{{\it RXTE}}
\newcommand{\ergcms}{erg cm$^{-2}$ s$^{-1}$}
\newcommand{\msun}{M$_{\odot}$}
\shortauthors{Muno et al.}
\shorttitle{Radio and X-ray Emission from \grsmq}

\begin{document}

\title{Radio Emission and the Timing Properties of the Hard X-ray State of
\grsmq}
\author{Michael P. Muno, Ronald A. Remillard, Edward H. Morgan,}
\affil{Department of Physics and Center for Space Research, Massachusetts 
Institute of Technology, 77 Massachusetts Ave., Cambridge, MA 02139}
\email{muno@mit.edu, ehm@space.mit.edu, rr@space.mit.edu}
\author{Elizabeth B. Waltman}
\affil{Remote Sensing Division, Naval Research Laboratory, 
Code 7210, Washington, DC 20375}
\email{ewaltman@rsd.nrl.navy.mil}
\author{Vivek Dhawan, Robert M. Hjellming}
\affil{National Radio Astronomy Observatory, Socorro, NM 87801}
\email{vdhawan@aoc.nrao.edu}
\author{Guy Pooley}
\affil{Mullard Radio Astronomy Observatory, Cavendish Laboratory,
Madingley Road, Cambridge CB3 0HE}	
\email{ggp1@cam.ac.uk}
\begin{abstract}

We combine a complete sample of 113 
pointed observations taken with the {\it Rossi X-ray Timing Explorer}
between 1996--1999,
monitoring observations taken with the Ryle telescope and the 
Green Bank Interferometer, and selected observations with the 
Very Large Array to study the radio and X-ray properties of 
\grsmq\ when its X-ray emission is hard and steady. We establish that radio 
emission always accompanies the hard-steady state of \grsmq, but
that the radio flux density at 15.2 GHz and the X-ray flux between
2--200 keV are not correlated. Therefore we study the X-ray spectral 
and timing properties of \grsmq\ using three approaches:
first, by describing in detail
the properties of three characteristic observations, then by displaying
the time evolution of the timing properties during periods of both faint 
and bright radio emission, and lastly by plotting the timing properties
as a function of the the radio flux density.
We find that as the radio emission becomes brighter and more optically
thick, 1) the frequency of a ubiquitous 
0.5--10 Hz QPO decreases, 2) the Fourier phase lags between hard 
(11.5--60 keV) and soft (2--4.3 keV) in the frequency range of 
0.01--10 Hz change sign from negative to positive, 3) the coherence between
hard and soft photons at 
low frequencies decreases, and 4) the relative amount of low 
frequency power in hard photons compared to soft 
photons decreases. We discuss how these results reflect upon
basic models from the
literature describing the accretion flow around black holes and the 
possible connection between Comptonizing electrons and compact 
radio jets.

\end{abstract}

\keywords{black hole physics--- stars: individual (GRS 1915+105) --- X-rays: stars}

\section{Introduction}

The microquasar \grsmq\ is one of the most interesting galactic sources 
of radio,
infrared, and X-ray emission. It was first discovered as a transient
X-ray source by {\it Granat} in 1992 \citep{ct92}, and later was 
observed to emit highly relativistic radio jets with intrinsic 
velocities greater than 0.9c \citep{mr94, fen99b}. High optical 
extinction has prevented studies of the binary companion of \grsmq, 
and there is no known 
orbital period or mass function for the system. However, the X-ray luminosity
of \grsmq\ greatly exceeds the Eddington luminosity for a neutron star 
(Greiner, Morgan, and Remillard 1996), and the other galactic sources
of highly relativistic ($> 0.9$c) jets, \gromq\ \citep{ob97, sha00} and
V4641~Sgr \citep{oro01}, have mass functions greater 
than the maximum mass for a neutron star, 3 \msun. \grsmq\ is therefore 
thought to be a black hole binary system. The properties of microquasars
in the galaxy are reviewed by \citet{mr99}.

Although \grsmq\ is often noted for exhibiting X-ray, radio, and 
infrared emission which is variable on time scales of seconds to minutes
\citep{gmr97, pf97, mir98, eik98},
the source also exhibits steady emission which is reminiscent of 
black hole candidates such as 
\cygxone\ and \gxbh. Both the steady and variable emission 
from \grsmq\ exhibit two (Muno, Morgan, \& Remillard 
1999; Rao, Yadav, \& Paul 2000) 
or three \citep{bel00} basic modes of X-ray emission. 
In the two-state description, the hard state resembles the very high state of 
canonical black hole binaries (Morgan, Remillard, \& Greiner 
1997),
as it exhibits X-ray flux with a power-law spectrum above 40 keV, 
prominent thermal emission from the accretion disk below $\sim 5$~keV, and a
strong (up to 15\% RMS) 0.5--10 Hz quasi-periodic oscillation (QPO).
Conversely, the soft state of \grsmq\ exhibits little X-ray emission 
above 40 keV and no 0.5--10 Hz QPO, although a stationary 67 Hz QPO 
and several low frequency ($< 0.1$ Hz) QPOs are often observed during this
state. 

Long-term radio and X-ray monitoring of \grsmq\ 
\citep{rod95, fos96, har97, pf97, ban98} has revealed 
that periods of hard, steady X-ray emission are often accompanied by 
bright, optically thick radio emission. This has been dubbed 
the ``plateau state'' by Foster et al. (1996; see also Fender 2001).
The radio plateau emission in \grsmq\ resembles steady, optically thick 
radio emission observed coincident with hard X-ray emission in the 
black hole candidates \cygxone\ \citep{hgo75, bro99}, 
\gxbh\ \citep{cor00}, and \cygxthree\ \citep{wal95, mcc99}. 
The connection between the X-ray and radio emission 
in \grsmq\ is illustrated in Figure~\ref{asm}, where we 
plot the intensity as a function of time as observed in soft X-rays 
by the All-Sky Monitor (ASM) aboard the {\it Rossi X-ray 
Timing Explorer} (\rxte; first panel), and at radio wavelengths by 
the Ryle telescope (15.2 GHz; third panel). We also plot the 
the ASM hardness ratio HR2 (5-12 keV / 3-5 keV; second panel),
and the radio spectral index $\alpha$ taken from monitoring data 
at 2.25	and 8.3 GHz from the Green Bank Interferometer (i.e. 
$S_{\nu} \propto \nu^{\alpha}$; only observations where $S_\nu$ at
both wavelengths are greater than 20 mJy are plotted). 
Bright radio plateau emission and hard-steady X-ray emission is evident 
on several occasions 
(e.g. MJD~50730--50750), yet the radio emission is sometimes a factor
of 10 fainter during other hard X-ray states (e.g. MJD~50450--50550). 
Observations with the Very Long Baseline Array during
the plateau state reveal that the optically thick radio 
emission originates in a compact jet on order 10 AU across 
(Dhawan, Mirabel, \& Rodr\'{\i}guez 2000). 

\placefigure{asm}

The radio plateau state is of further interest because it is often preceded
and/or followed by optically thin radio flares in \grsmq\ \citep{fen99b}, 
as are similar hard-steady states with optically thick radio emission in 
\gxbh\ \citep{cor00} and  \cygxthree\ \citep{mcc99}.
The optically thin emission in \grsmq\ originates from relativistic ejections
of synchrotron-emitting blobs detached from the central source 
\citep{mr94, fen99b, dmr00}. In Figure~\ref{asm}, optically
thin radio flares precede and terminate the radio plateau emission which 
occurred between MJD~50730--50750, while weaker flares with shorter duration 
terminate the long period of weaker radio emission from 
MJD~50450--50550. These events are associated with flares in the ASM 
light curve, which suggests that the most powerful jets are associated with 
instabilities in the inner accretion disk. Moreover, there have been 
several occasions in which weaker radio and infrared flares
recur at $\sim$30 minute intervals, and are preceded by distinctive dip 
and flare patterns in X-rays \citep{pf97, eik98, mir98}.

During the plateau state of \grsmq, the radio luminosity is low compared to
the large optically thin flares, but the power in the radio jets may still 
represent a significant fraction of the accretion energy inferred 
from the X-ray emission \citep{fb99, fen01}. Since this emission originates 
from a compact jet within 10 AU of the central source and is present only 
during the hard X-ray state, it seems very likely 
that the compact jet is influenced by the properties of 
the inner accretion flow, and vice versa. This has led 
\citet{fen99a} to suggest
that the hard X-rays originate from inverse Compton scattering of seed
photons at the base of the compact jet of relativistic electrons. 
This idea motivates further efforts to compare the properties
of the radio emission to the X-ray spectral and timing information during
the hard state of \grsmq. 

The timing properties of \grsmq\ provide a promising means of 
probing the accretion flow near the black hole. In particular, the 
0.5--10~Hz QPO observed during the hard state appears to provide 
a link between the optically thick accretion disk and the Comptonizing
electrons, because (1) the frequency of the QPO increases as the thermal 
flux from the disk increases (Chen, Taam \& Swank 1997; 
Trudolyubov, Churazov, \& Gilfanov 1999; Markwardt, Swank, \& 
Taam 1999), 
(2) the spectrum of the QPO amplitude is hard, indicating an origin 
in the power law component 
of the spectrum \citep{mrg97}, and (3) the QPO is only present when the 
power-law component of the X-ray spectrum is strong, while it is absent 
when the power-law component is weak \citep{mmr99}. Additional timing 
characteristics, such as the Fourier 
phase lags and the coherence function, may further constrain the relationship 
between hard and soft X-ray components. \citet{rei00} 
have found that the signs of the phase lags change from positive to negative 
as the frequency of the 0.5--10 Hz QPO increases, which has provided impetus
for others \citep[e.g.][]{nob00} 
to develop models describing the structure of the Comptonizing 
corona and how it is related to the optically thick accretion disk.

In this paper, we use a complete sample of ``hard-steady'' observations 
of \grsmq\ (referred to as the $\chi$ state by Belloni et al. 2000) 
to examine in detail the connection between the hard,
steady X-ray emission and the steady radio emission. First,
we establish that the hard-steady X-ray states of \grsmq\ always 
exhibit detectable radio emission. However, the intensities of the 
2--200~keV X-ray and 15.2 GHz radio flux are not correlated, 
so we proceed to examine
whether there are more subtle relationships between the spectrum and
timing properties of the X-ray emission and the intensity and spectrum 
of the radio emission.  We investigate these 
correlations in three different manners: first, by describing in detail
the properties of three characteristic observations, then by displaying
the time evolution of the timing properties during periods of both faint 
and bright radio emission, and lastly by examining the timing properties
as a function of the radio flux density.
  
\section{Observations and Data Analysis}

\placetable{rxte}

\subsection{Ryle Monitoring Observations}

The Ryle Telescope at the Mullard Radio Astronomy Observatory, Cambridge, is
primarily used for microwave-background measurements, but during gaps
in that program it has been used to monitor variable galactic
and extra-galactic sources. It operates at 15.2 GHz, and uses a single
set of linearly-polarized feeds. 
In good observing conditions, the RMS noise in a 1 minute integration 
is about 3.5 mJy. Since the measurements are unbiased, the noise level 
decreases as the square root of the observing time. 
Details of the observing and analysis routine are given in \citet{pf97}.

We have averaged the radio measurements in two ways for the purposes
of this paper. We have used the averaged daily flux density measured 
with the Ryle Telescope when displaying monitoring light curves of 
\grsmq\ (e.g. Figure~\ref{asm}). When comparing the Ryle measurements to
the properties of pointed \rxte\ observations, we have averaged data 
taken within 0.5 days before and after the \rxte\ observation in order
to sample the radio properties which most closely correspond to the X-ray
observation (see Table~\ref{rxte}). 

\subsection{Green Bank Interferometer Monitoring Observations}

Monitoring with the Green Bank Interferometer has been performed 
simultaneously at 
2.25 and 8.3 GHz on a 2.4 km baseline with 35 MHz bandwidth and dual circular
polarization.  Most scans are 10-15 minute integrations, and the calibration
procedure is described by \citet{fos96}.  Random (one sigma) 
errors in the GBI data are flux dependent: 4 mJy (2 GHz) or 6 mJy (8 GHz) 
for fluxes $< 100$ mJy, 15 mJy (2 GHz) or 50 mJy (8 GHz) for fluxes $\simeq 1$ 
Jy.  In addition, systematic errors are introduced by atmospheric and 
hardware effects, which we estimate may approach 10\% at 2 GHz and 20\% at 
8 GHz, occasionally higher for extreme local hour angles. Flux density
measurements below 20 mJy approach the noise level of the instruments, and
therefore we do not include measurements below this level in our analysis.

As with the Ryle data, daily GBI measurements have been averaged when producing
monitoring light curves, and GBI data taken within 0.5 days of a pointed 
\rxte\ observation have been averaged when comparing individual observations
(Table~\ref{rxte}).
In both cases, we also have computed the spectral index $\alpha$
from the average flux in the 2.25 and 8.3 GHz bands.

\subsection{Very Large Array Observations}

The VLA is a multi-frequency, multi-configuration aperture
synthesis imaging instrument, consisting of 27 antennas of 25 m
diameter. The receivers at 1.42, 5.0, 8.45, and 15.2 GHz, have been used on
various occasions, with 2 adjacent bands of 50 MHz nominal bandwidth
processed in continuum mode.  The corresponding 1-sigma sensitivities
in 10 minutes  are 0.06, 0.06, 0.05, and 0.2 mJy respectively.
Switching between multiple frequencies takes about 30 s. The array
configuration is varied every 4 months to cycle between 4 sets,
A, B, C, and D,  with maximum baselines of about 36, 11, 3.4, and 1 km.

Observations of \grsmq\ while in its radio faint state have been obtained 
from the public archive in order to constrain the radio spectrum. 
The observations we have found are listed in 
Table~\ref{vla}, along with mean fluxes from Ryle and GBI observations
taken within a half day of the VLA observations.
For all of the observations reported here, \grsmq\ was unresolved
by the synthesized beam (see Table~\ref{vla})
 in any configuration at any wavelength. Other sources in the field of
view (mainly at 1.4 GHz) are well separated in the images, so there is
no confusion with \grsmq.

\placetable{vla}

Calibration and imaging were carried out with standard tasks in the
NRAO AIPS (Astronomical Image Processing System) package. For all of
these observations, the phase calibrator was 1925$+$211. The primary flux 
density calibrator was 3C286 (1328$+$307) for all observations except
1996 November 14, when 3C48 (0134$+$329) was used. In practice, the flux
density errors are not set by the RMS receiver (thermal) noise stated
above, but by errors in the flux density scale, estimated to be 3-5\% of
the measurement, and/or source variability, depending on the
occasion. More details of the VLA are given in the Observational Status 
Summary
(\url{http://www.aoc.nrao.edu/vla/obstatus/vlas/vlas.html}), and the data 
reduction is treated in formal detail in Taylor, Carilli, \& Perley
(1999) and in the VLA cookbook (\url{http://www.cv.nrao.edu/aips/cook.html}).

\subsection{Pointed \rxte\ Observations}

\begin{figure}
\plotone{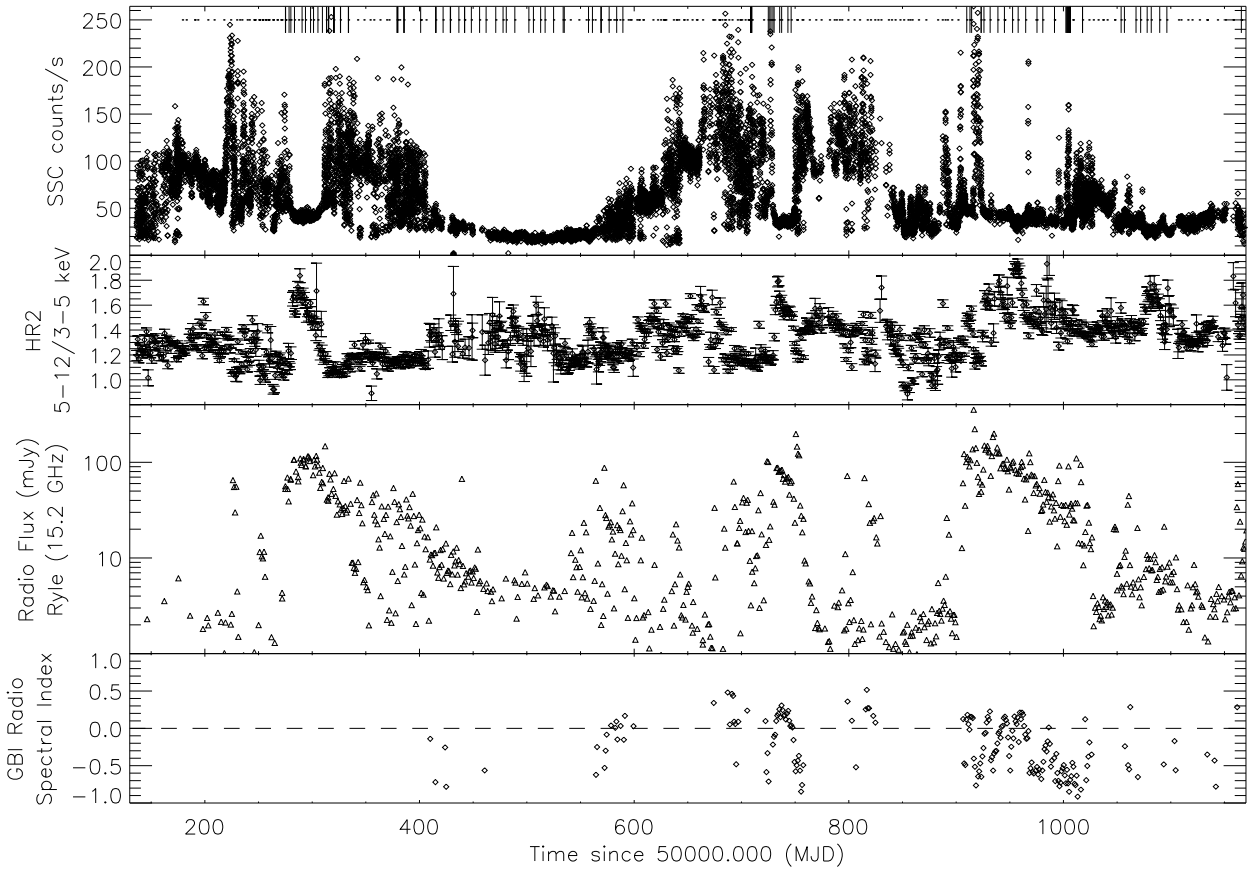}
\caption{Monitoring data from X-ray and radio wavelengths. 
{\it Top panel}: Count rate as a function of time from the \rxte\ ASM,
1.5--12 keV. Vertical bars across the top represent hard-steady 
pointed PCA/HEXTE observations. Dots represent the remainder of the 
pointed observations.
{\it Second panel}: \rxte\ ASM hardness ratio 
HR2 (5-12 keV / 3-5 keV). {\it Third panel}: Flux as a function of 
time from the Ryle telescope at 15.2 GHz. 
{\it Bottom panel}: GBI radio spectral 
index ($\alpha = \Delta \log S_\nu / \Delta \log \nu$, where 
$\Delta \log S_\nu$ is the difference between the logarithm of the 
mean daily flux at $\nu =$ 8.3 and 2.25 GHz; points are plotted only if
the flux density at both frequencies was larger than 20 mJy).}
\label{asm}
\end{figure}

The {\it Rossi X-ray Timing Explorer} (\rxte) consists of three instruments:
the All-Sky Monitor \citep[ASM;][]{lev96}, the Proportional Counter Array
\citep[PCA;][]{jah96}, and the High Energy Timing Experiment 
\citep[HEXTE;][]{roth98}.
Figure~\ref{asm} shows the ASM light curve of \grsmq\ along with daily radio
measurements from Ryle and the spectral index from the GBI. There
is no apparent correlation between the radio emission and the ASM count rate,
although there is some correlation with the ASM hardness ratio HR2 
(5-12 keV / 3-5 keV). In addition, there has been a long campaign of
weekly observations with with the PCA and HEXTE, which we use to study
the detailed timing and spectral properties of the hard emission from 
\grsmq. We define intervals
as ``hard-steady'' if the ratio of the mean PCA count rates 
in the 12--60 keV to the 2--12 keV bands exceeds 0.05 and 
the mean variability in the count rate (2--60 keV) at 1s time intervals 
is less than 12\% for an \rxte\ orbit (typically 3000~s). 
In Figure~\ref{asm} we have marked the pointed \rxte\ observations of the 
hard-steady state with vertical bars above the ASM light curve, and 
we have listed them in Table~\ref{rxte}.
The small dots represent the remainder of the pointed observations.
We have analyzed all 113 observations of \grsmq\ in the hard steady 
state taken with \rxte\ between 1996 April 15 (MJD~50188) 
through 1999 March 22 (MJD~51259), during the time interval for the PCA 
instrument gain setting known as ``epoch 3''. In order to improve the 
signal-to-noise of the
spectral and timing measurements whenever possible, if the variability
in the mean count rate in 16 s bins is less than
5\% during a single day, we have analyzed all of the data for that day
together. Otherwise, we have 
grouped the data while remaining within the 5\% variability limit (16 s bins), 
with the smallest unit being a single \rxte\ orbit.

For each time unit, we have integrated 128-channel energy spectra 
(``Standard2'') data from PCUs 0 and 1 of the PCA, and 64 channel 
(archive mode) spectra from both clusters of HEXTE. In order to calculate the
flux from GRS 1915+105, we have fit the spectra with the phenomenological 
model described in \citet{mmr99}. The model consists of 
the sum of a multi-color blackbody, a power law (with a high-energy
cutoff when necessary), and a Gaussian line (representing iron emission 
between 5--7 keV). Interstellar absorption is taken into account assuming
a column density equivalent to $6\times10^{22}$ cm$^2$. A 
normalization factor is allowed to account for differences in the effective 
area of the PCUs and the HEXTE clusters which are not taken into 
account by the response matrices. The standard FTOOLS 4.2 background 
subtraction algorithm for bright sources has been applied to the data, 
and a systematic uncertainty of 1\% has been added to each spectral bin. 
We have computed an X-ray flux for each 
observation, by summing the bolometric flux implied by multi-temperature
disk model ($F_{bb} = 1.08\times 10^{-11} N_{\rm app} \sigma T_{\rm app}^4$),
the flux in the power law between 2--200 keV 
($\int_2^{200} 1.60 \times 10^{-9} N_\Gamma E^{-\Gamma+1}\exp[-E/E_{\rm cut}] dE$),
and the flux in the iron line 
($N_{\rm Gauss}/(\sqrt{2\pi}W_{\rm Gauss})\int_0^\infty \exp[ 0.5(E - E_{\rm Gauss})^2/W_{\rm Gauss}^2 ] dE$).

We also have created power density spectra (PDS) and cross spectra (CS) 
for each 256 s interval of data with $2^{-7}$ s time resolution and four 
energy channels, using combinations of the binned and event modes for 
each observation. The four energy channels used are as follows: 
2--4.3 keV (henceforth referred to as the ``low energy'' band), 4.3--7.8 keV, 
7.8--11.5 keV, and 11.5--60 keV (referred to as the ``high energy'' band). 
The PDS and CS have been averaged for the entire time unit (an \rxte\ orbit or 
an observation, as defined above) and logarithmically re-binned. 
The PDS have been corrected for dead-time effects and Poisson noise 
\citep{mrg97}, and have been normalized to the fractional RMS squared per Hz. 
To quantify the features of the PDS, we have fit PDS from each energy 
band with a model consisting of a power law with two breaks and Lorentzians 
for any QPOs present in the PDS. We have searched for the 67 Hz QPO
during each of these hard-steady observations using data with $2^{-13}$ s time
resolution in a single energy channel, but we did not detect it. Typical
upper limits range from 0.1--0.5\% (1 $\sigma$).

The coherence function and phase lags have been calculated from the CS in 
the manner described in \citet{vn97}. We have used the standard convention
that ``hard'' or ``positive'' lags indicate that higher energy photons 
lag behind lower energy photons.
The uncertainties have been calculated for the
case of high signal power and high measured coherence, using an estimate
of the dead-time-corrected Poisson noise \citep{mrg97}. 
If the signal power is smaller than the estimated noise, or if 
the coherence is less than the noise divided by the signal, 
we do not plot either the phase lags or the coherence functions.

Guided by the long-term monitoring light curves in Figure~\ref{asm}, 
we have defined subsets of the hard-steady observations based upon the radio
emission from \grsmq\ in order to facilitate the comparison between the
radio and X-ray emission. We have defined ``radio plateau'' hard-steady 
conditions for observations for which \grsmq\ is brighter than
20 mJy at 15.2 GHz, and exhibits a flat radio spectrum with index (measured
from the daily average of the flux) $\alpha > -0.2$. We also have 
defined ``radio
steep'' conditions for observations when \grsmq\ is brighter than
20 mJy at 15.2 GHz, and exhibits a optically thin radio spectrum with index 
$\alpha < -0.2$. Figure~\ref{asm} demonstrates that the radio steep conditions 
typically represent the transition into and out of periods of radio 
plateau emission,
and occur during the decays of large optically thin radio flares 
\citep[see also][]{fen99b, dmr00}. 
Our discussion of radio steep observations will be limited, as the optically 
thin emission probably results from radio ejecta that could evolve 
independently of
the instantaneous X-ray conditions on 0.5 day time scales.

Finally, we have defined ``radio-faint'' conditions for 
those hard-steady X-ray observations for which the daily 
average of the radio flux at 15.2 GHz is less than 20 mJy, because for these 
observations the radio flux at the GBI frequencies is too low to reliably
measure a spectral index. 
The radio faint conditions occur with two variants--- those observations 
for which the X-ray emission is bright, and those for which it is faint.
The behavior of the timing properties with respect to the radio emission
is similar for both variants, although
the shape of the X-ray spectrum and the values of the timing properties
(e.g. the QPO frequency and the phase lags) are distinctive, 
and warrant a separate discussion. 

\section{Results}

\begin{figure}
\plotone{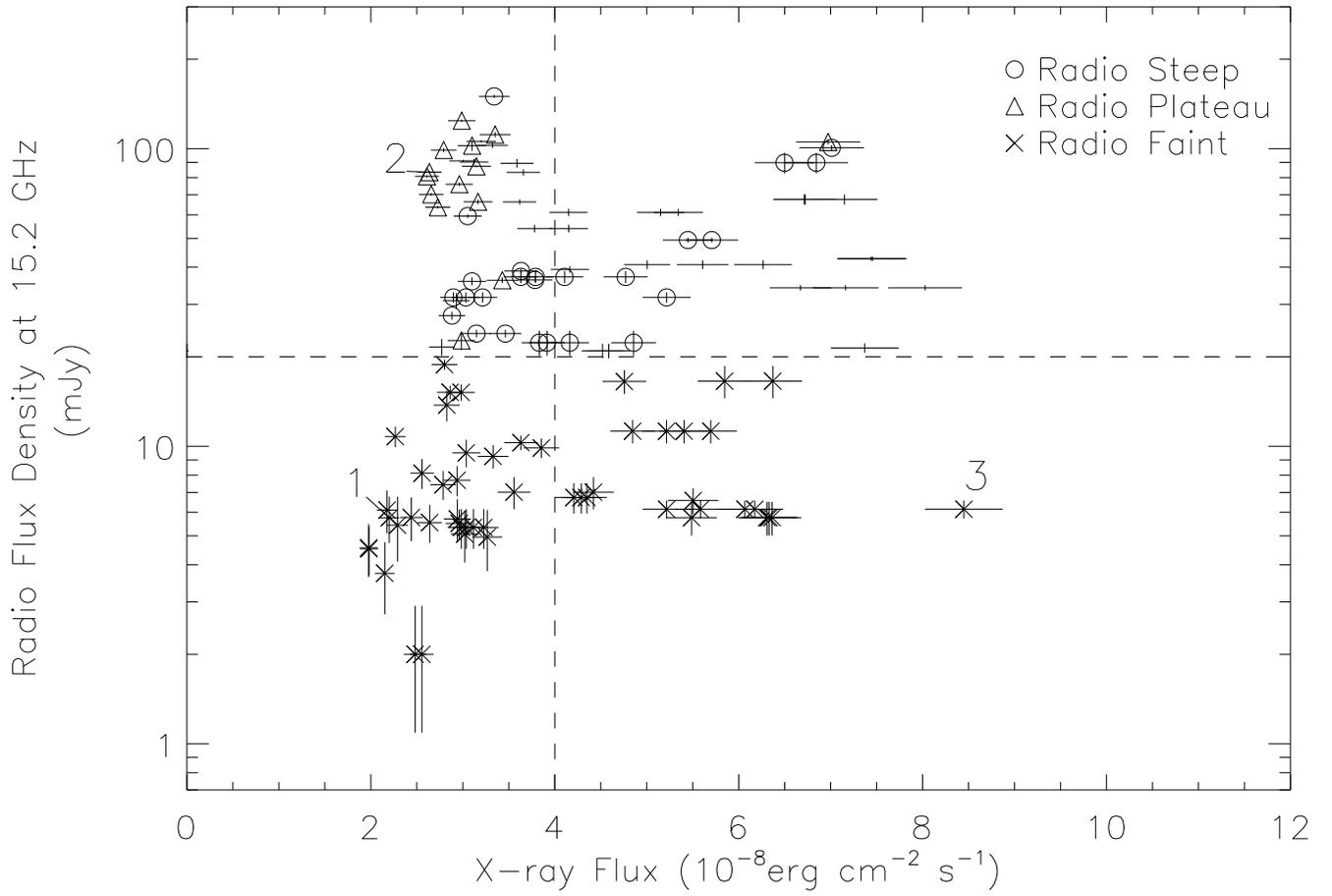}
\caption{The radio flux density at 15.2 GHz measured with the 
Ryle telescope as a function of the 2--200 keV X-ray flux measured 
with the PCA 
aboard \rxte. All hard-steady observations exhibit detectable radio 
emission, although the strength of the X-ray and radio emission are
not correlated. The radio conditions of each observation are indicated
as follows: triangles represent radio-plateau 
hard-steady conditions, circles radio-steep conditions, and $\times$'s
radio-faint conditions. Three characteristic observations are indicated 
in the diagram: (1) a radio-faint, X-ray faint observation, (2) a radio
plateau, X-ray faint observation, and (3) a radio-faint, X-ray bright
observation.}
\label{rvx}
\end{figure}

We first investigate 1) whether all of the hard-steady observations 
are coincident with radio emission, 2) whether the X-ray and 
radio flux are correlated, and 3) how the radio conditions
(plateau, steep, and faint) are distributed in a plot of radio 
vs. X-ray flux.  Figure~\ref{rvx} displays a plot of the radio flux 
density at 15.2 GHz (log scale) versus the 2--200~keV
X-ray flux for each of the hard-steady observations. It is clear that 
all hard steady observations have radio flux densities greater than 2 mJy.
Panel 3 of Figure~\ref{asm} shows that the radio flux 
from \grsmq\ does drop below 1 mJy (e.g. MJD~50825--50875), but such 
intervals are associated with the soft X-ray state. Thus, {\it radio
emission always accompanies the hard-steady X-ray state}, although 
a note of caution is appropriate for those observations associated with 
variable, optically thin radio emission, as noted above.
However, Figure~2 also shows the radio and X-ray fluxes are not correlated
in the hard-steady state. Likewise, we have found that the total
50-100 keV flux, the power law flux, and the thermal (multi-color disk) flux
are not correlated with the strength of the radio emission (not shown). 

\placetable{rvx}

The radio conditions are represented in Figure~\ref{rvx} as follows: 
radio plateau observations are plotted as triangles, radio steep
as circles, and radio faint as $\times$'s. Observations for which the GBI
spectral index has not been measured are plotted only with error bars.
The plot can be roughly split into four quadrants. The
radio plateau observations are clearly localized in the upper left 
quadrant, with $S_{15.2} > 20$ mJy and $F_{\rm X} < 4\times 10^{-8}$ \ergcms.
Some radio steep observations are also present in this quadrant.
The upper right quadrant contains radio steep observations and observations
for which no spectral index is available from the GBI. These latter
observations were taken between MJD~50313--50320 (before the current GBI 
monitoring program), a time interval which is included in the analysis of 
\citet{ban98}. During this time span a large radio flare
occurred terminating a plateau state, and the radio flux density
at 1.4 GHz was consistently larger than that at 3.3 GHz, which 
suggests that these points should be classified as radio steep. 
Therefore, the upper right quadrant ($S_{15.2} > 20$ mJy and 
$F_{\rm X} > 4\times 10^{-8}$ \ergcms) appears to include only radio steep 
observations, during which the radio and X-ray emission may vary 
independently since the radio measurements may be dominated by the 
evolution of an expanding jet detached from the central source. 
For this reason, we choose to ignore the upper right quadrant. 

The lower quadrants, with $S_{15.2} < 20$ mJy, contain radio faint
observations by construction. After examining the temporal clustering of these
radio-faint observations, we feel that it is useful also
to consider them in two quadrants: an X-ray faint
quadrant with $F_{\rm X} < 4\times 10^{-8}$ \ergcms, and an X-ray bright
quadrant with $F_{\rm X} > 4\times 10^{-8}$ \ergcms. This separation is 
further motivated by their differing energy spectral and timing 
properties, which are described in the next section.

\subsection{Characteristic Observations}

From each of the three quadrants under consideration from the radio versus 
X-ray flux plane (Figure~\ref{rvx}), we 
have selected a characteristic observation in order to illustrate the
range of X-ray spectral and timing properties within the hard-steady
X-ray state. In Figures~\ref{spec}, \ref{pdsetc}, and \ref{vsspec},
the left panel is a radio faint, X-ray faint observation (MJD~50488; 
1997 February 9);
the center panel is a radio plateau, X-ray faint observation (MJD~50737;
1997 October 16);
and the right panel is a radio faint, X-ray bright observation (MJD~50708;
1997 September 17). The radio and X-ray fluxes from these three 
observations are indicated in Figure~\ref{rvx}.

\subsubsection{X-ray Energy Spectra}

\begin{figure}
\plotone{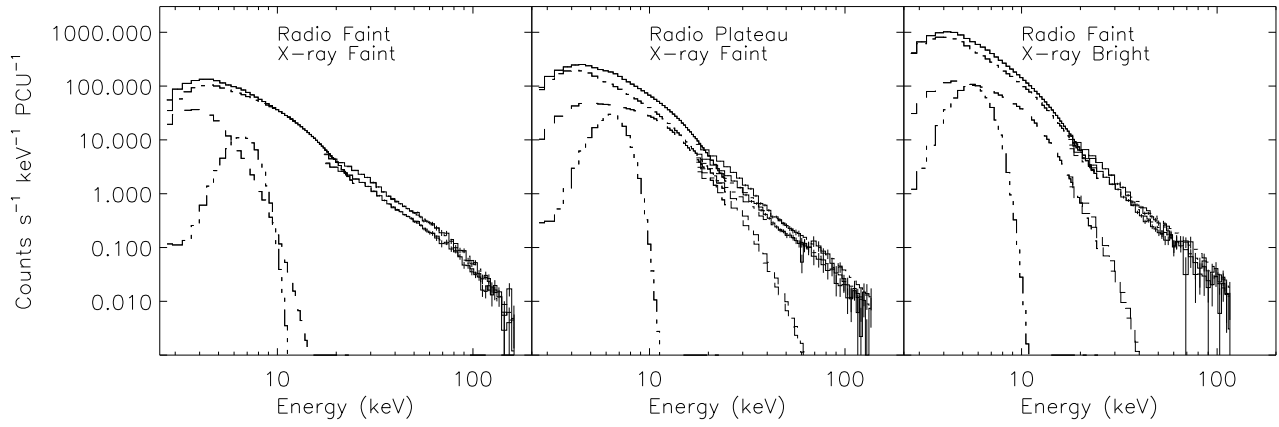}
\caption{The energy spectrum and the corresponding spectral
fits, as described in the text, for three characteristic observations:
a radio-faint and X-ray faint observation on MJD~50488, a radio-plateau 
observation on MJD~50737, and a radio-faint and X-ray bright observation
on MJD~50708. The power law component dominates
the spectrum at all energies, while the multi-temperature disk and
the Gaussian iron line contribute a smaller fraction of the flux.}
\label{spec}
\end{figure}

\placefigure{spec}

Figure~\ref{spec} illustrates the results of the spectral fits for 
the three representative observations 
with differing values of the radio and X-ray flux. 
The spectrum in the left panel is from a radio faint observation 
when the X-ray emission was also relatively faint 
($2.2\times10^{-8}$\ergcms). The soft emission during this observation 
can be modeled ($\chi^2_\nu \sim 1$) using a cool multi-temperature
blackbody with an apparent 
inner disk temperature of $T_{\rm app} \simeq 0.8$ keV and an apparent radius of 
$R_{\rm app} \simeq$40--80 km (assuming a distance of 11 kpc and an 
inclination angle 
of 66$^\circ$; see Fender et al. 1999b). 
The power law has a photon index $\Gamma \simeq 2$ and 
requires an exponential cutoff at about 70 keV. The results of this spectral
fit are characteristic of observations during which both the X-ray and radio
flux were low.

The spectrum in the middle panel is from an observation which exhibited
radio plateau conditions when the X-ray emission was also faint 
($2.6 \times10^{-8}$ \ergcms). Despite the similarity in X-ray flux to the
left panel, the shape of the X-ray spectrum is quite different.
The thermal component of the model
appears hot ($T_{\rm app} > 3$ keV) with a 
small apparent inner radius ($R_{\rm app} < 5$ km). 
This quasi-thermal component is not consistent with simple models of 
reflection from an accretion disk \citep[e.g.][]{mz95}. It is possible
that this broad thermal bump represents emission from a relatively 
small area of the disk which has been heated by, for example, a magnetic
flare (di Matteo, Celotti, \& Fabian 1999) or a spiral shock wave 
\citep{tp99}, or that the 
disk develops an optically thick atmosphere which 
Comptonizes the emergent spectrum, so that the spectral hardening and 
decreased apparent radius are illusory effects of increased scattering in a
``puffed up'' disk (Merloni, Fabian, \& Ross 2000). 
The photon index of the power law is $\Gamma \simeq 2.7$, and a cut-off 
at about 80 keV is required for observations with longer integration 
times and hence better statistics at energies above 100 keV. 
We note that \citet{rao00} have presented a different model
for spectra of \grsmq\ such as these, consisting of a multi-temperature
disk, a thermal-Compton spectrum \citep{st80}, and a power law. 
For the purposes
of this paper, we prefer to use our simpler model with fewer free parameters.

The right panel is from an observation
when the radio emission was faint (6 mJy) and the X-ray emission 
was bright ($8.4\times10^{-8}$ \ergcms). The spectrum is similar to 
the radio bright, X-ray faint observation in the center panel, except that
the relative contribution of the power law flux increases dramatically.
The thermal component is hot ($T_{\rm app} > 2.5$ keV) and has a small apparent
inner radius ($R_{\rm app} < 10$ km), while the power law index is steep 
($\Gamma \simeq$ 3.2).
The shape of the energy spectrum on 1997 September 17 is characteristic of all
X-ray spectra from the hard state of \grsmq\ when the total flux
is greater than $\simeq 4\times10^{-8}$ \ergcms.
The count rate from \grsmq\ tends to vary on time scales of 
hours when the source is bright in X-rays, so we only integrate spectra for 
individual \rxte\ orbits. With these exposure times, we can measure the 
power law only up to about 100 keV. However, when we combine 
spectra with similar photon indices, we
find that the data are consistent with no cut off in the power law below
200 keV.

The changes in the X-ray spectrum clearly do not track the radio flux from
\grsmq\ in a monotonic manner. The power law component steepens dramatically
as the radio emission increases in the X-ray faint state (left and center
panels of Figure~\ref{spec}), but is steepest when the X-ray emission is 
bright and the radio emission is faint (right panel). At low X-ray fluxes 
($< 4\times10^{-8}$ \ergcms), the thermal 
component is hot and has a small apparent inner radius during radio plateau 
observations, while it is cool and has a large apparent inner radius when
the radio emission is faint \citep[see also][]{mmr99}. 

\subsubsection{Power Density Spectra and Cross Spectra}

\begin{figure}
\plotone{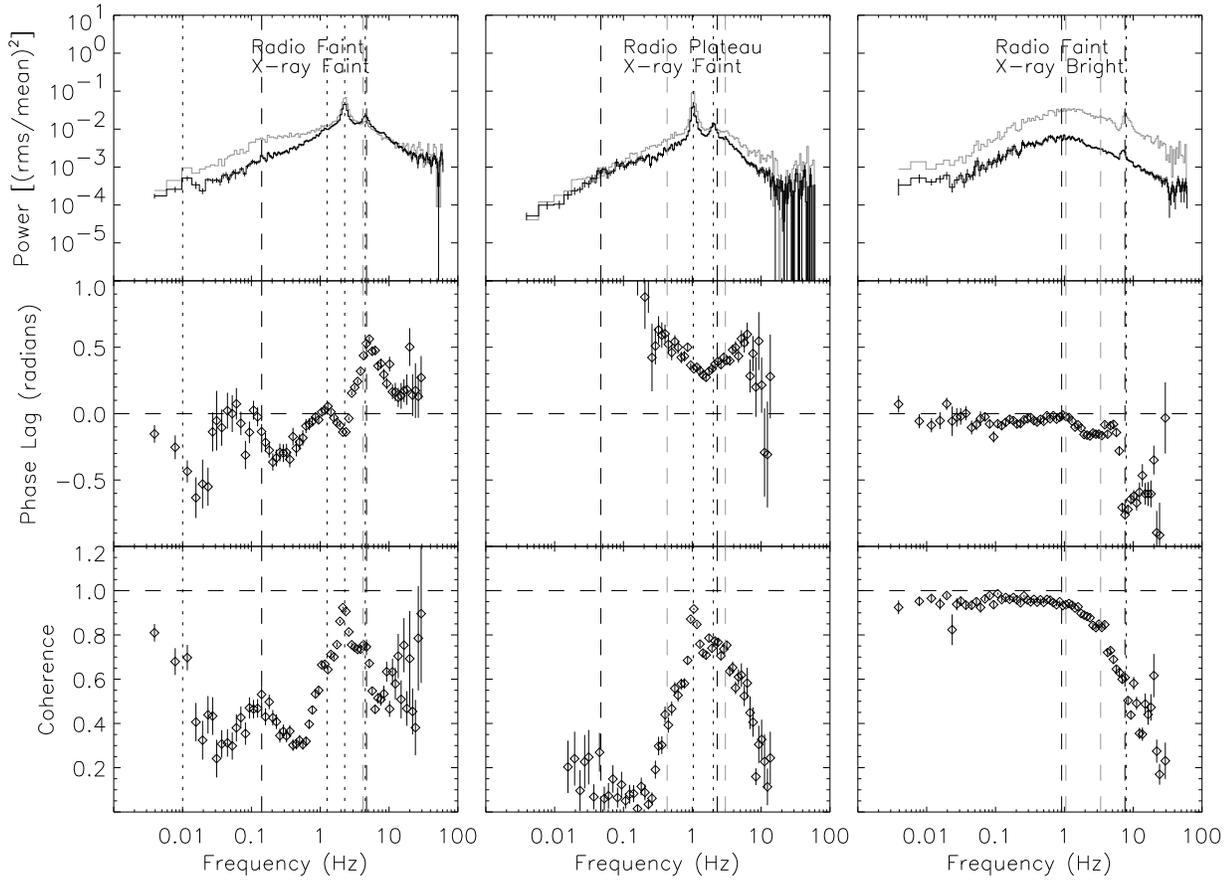}
\caption{Timing properties of the three characteristic observations in 
Figure~3. {\it Top panel}: PDS in two energy bands, the black line 
indicates the 2--4.3 keV energy band, while the  grey line 
indicates the 11.5-60 keV band. {\it Middle panel}: the phase lags 
between the 2-4.3~keV and the 11.5--60~keV bands. {\it Bottom panel}:
the coherence function between the 2-4.3~keV and the 11.5--60~keV bands.
In all panels, we have indicated the frequencies of 
the QPOs in our fits to the 2.5--5 keV band with black short-dashed 
lines, and the break frequencies with black long-dashed lines. We have 
also indicated the break frequency in the 11.5-60 keV band with grey 
long-dashed lines (see below).}
\label{pdsetc}
\end{figure}

\placefigure{pdsetc}

The power spectra (here displayed as the frequency times the power density), 
Fourier phase lags, and coherence functions of the three observations 
from Figure~\ref{spec} are displayed in Figure~\ref{pdsetc}. 
In the top panel, two lines are plotted for each power spectrum: 
the black line 
indicates the 2--4.3 keV (soft) energy band, while the the grey line 
indicates the 11.5--60 keV (hard) band. The PDS in all three examples of 
Figure 
\ref{pdsetc} have relatively flat spectra with a low frequency break, a QPO 
between 0.5--10 Hz, and a dramatic break at frequencies higher than the QPO. 
In the middle panel of 
Figure~\ref{pdsetc}, we plot the phase lags between 
the 11.5--60~keV and the 2--4.3~keV bands, and in the bottom panel we plot the 
coherence function for the same two energy bands. To simplify 
the comparison between the power spectra, 
the phase lags, and the coherence functions, 
we have indicated the frequencies of the QPOs in our fits to the 
2--4.3 keV band with vertical black short-dashed lines, and 
the continuum break frequencies with vertical black long-dashed lines. 
We have also indicated the 
break frequency in the 11.5--60~keV band with grey long-dashed lines 
(see below). For the following discussion we define three frequency
regions related for the features of the power spectrum: region~1 represents
frequencies lower than the low frequency continuum break, region~2 frequencies
between the low frequency break and the QPO frequency (less its width),
and region~3 frequencies higher than the QPO frequency (or its harmonic
if present).

The first point of interest is that the QPO frequency 
is not strictly correlated with the X-ray flux from \grsmq. From
the spectra in Figure~\ref{spec}, 
it is clear that the X-ray flux increases from 
left to right, while in Figure~\ref{pdsetc} the QPO frequency is lowest in 
the center panel (radio plateau conditions). In the following sections, we will
demonstrate that the QPO frequency is always lowest during radio 
plateau conditions.

The second and most obvious difference between the timing properties 
of these three observations is in the Fourier phase lags. In region~2 
(between the low frequency break and the QPO), the phase lags are
negative when the radio emission is faint (left and right panels), 
and positive during radio plateau emission 
\citep[center panels; compare][]{rei00}.
This behavior is in sharp contrast to sources such as \gxbh\ and \cygxone, 
for which the phase lags are always positive. Only the phase lags during the
radio plateau state resemble those of \gxbh\ \citep{nwd99} and 
\cygxone\ \citep{cui97}. 
 
The third notable result concerns the coherence 
function in region~1 (i.e. frequencies lower than the first break). 
The coherence function between the 11.5--60 keV and
the 2--4.3 keV bands is lowest ($\simeq 0$) during the radio plateau conditions 
(center panel), somewhat higher 
($\simeq 0.4$) when both the radio and X-ray emission are faint (left panel), 
and nearly unity when the radio emission is faint and the X-ray emission
is bright (right panel). Only the latter case
is reminiscent of \cygxone\ in its hard state \citep{cui97, now99}. 

Finally, a comparison of the amplitude and strength of the 
power spectra as a function of energy reveals further differences in 
each of the three observations. While the radio emission is faint, 
(right and left panels) the root mean power at low frequencies is greater 
in the 11.5--60 keV energy band (gray curve) than in the 2--4.3 keV band
(black curve). However, 
during radio plateau conditions (center panel) the root mean power is 
nearly equal in all energy bands. The shape of the power spectrum 
also changes--- the low frequency break is strongest in the low energy 
band in the radio plateau observation (center panel), while the break is 
most apparent in the high energy band in the radio faint, X-ray faint 
observation (left panel). When the X-ray emission is bright, 
the low frequency break is evident in both energy bands (right panel). 

There are several other interesting details in the timing 
properties presented in Figure~\ref{pdsetc}, although 
they are not directly relevant to the question of how the timing
properties are related to the radio emission. 
The harmonic of the QPO tends to be 
strongest ($\simeq 10$\%) when count rate is lowest (left and center
panels), but is not apparent during X-ray bright observations (right panel).
A weak feature at approximately
half of frequency of the strongest QPO peak is also 
evident during observations when both the X-ray and radio flux are 
faint (right panel), with an RMS amplitude of about $\simeq$1\% 
and $Q$ values of 1--10. 

Finally, a weak QPO is evident between 0.01--0.04 Hz during a stretch
of radio faint observations from MJD~50450--40560 (left panel), with an RMS 
amplitude of about 0.1\% in the 2-4.3 keV energy band. This low frequency 
QPO disappears at higher energies. Its centroid frequency decreases
as the frequency of the 0.5--10 Hz QPO decreases (not shown).
Since the 0.01--0.04 Hz QPO is present only in the low energy band, the 
coherence function at the QPO frequency is near zero.

\subsubsection{Relation of the Timing Properties to the Energy Spectrum}

The spectral fits which we have made to \grsmq\ indicate that 
two spectral components are present (Figure~\ref{spec}). 
Most models of the emission from X-ray binaries invoke 
inverse-Compton scattering to produce hard photons from soft 
seed photons, and consequently the variability in the hard and soft 
energy bands should be intimately related. However, before finalizing
any conclusions regarding the phase lag and coherence measures and how
these change with photon energy,  we need to examine how these
timing properties depend upon  
the amount of the flux which is contributed by each of the 
spectral components in the relevant energy bands.

\begin{figure}
\plotone{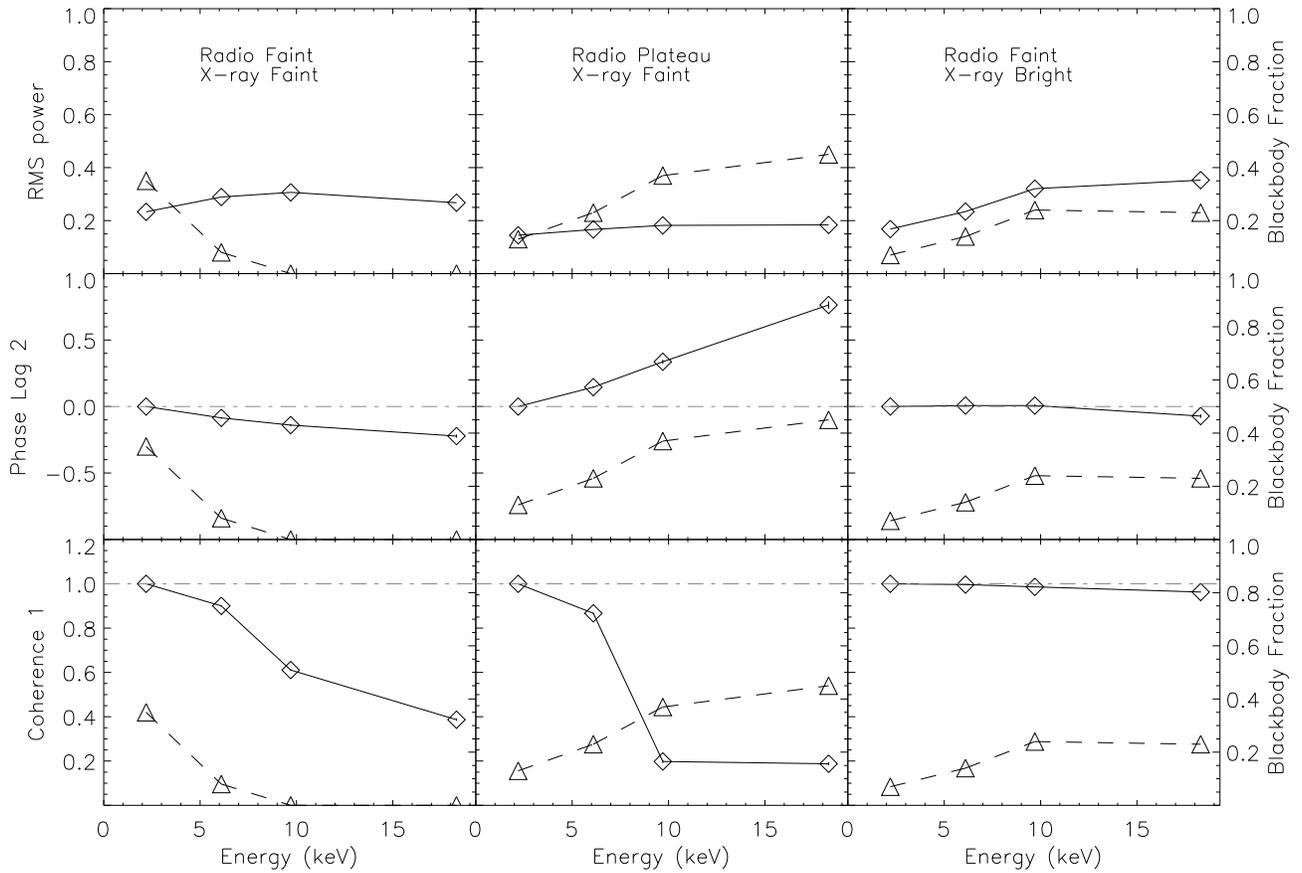}
\caption{The relation of the timing properties to the X-ray energy spectrum
for the three observations in Figure~3.
{\it Top panels}: The RMS variability at all frequencies as a function of 
energy plotted with diamonds connected by the solid line. 
{\it Middle panels}: The mean phase lag between the low frequency 
break and the QPO frequency 
(region~2) as a function of energy plotted with diamonds connected by 
the solid line. Positive values indicate that the
hard photons {\it lag} the soft photons. {\it Bottom panels}: The average
coherence function  
measured below the low frequency break (region~1) plotted with 
diamonds connected by the solid line. 
{\it All Panels}: Fraction of the count rate contributed by
the multi-temperature disk model (as derived from the spectral fits) as
a function of energy, plotted as triangles connected by the dashed line,
and labeled on the axes to the right.} 
\label{vsspec}
\end{figure}

\placefigure{vsspec}

In Figure~\ref{vsspec}, we examine the dependence of the timing properties
on photon energy for the same three representative observations as
in Figures~\ref{spec} and \ref{pdsetc} (diamonds connected by solid lines).
In all of the panels, the fraction of the count rate contributed by 
the disk component is displayed with triangles connected with dashed lines,
using the scale displayed on the right axis. Here it can be seen that
the thermal component contributes no more than half of the flux in all 
energy bands for the observations which we analyze in this 
paper.  Examining next the top panels, 
we find that the presence of the thermal component does not greatly affect the
RMS power in a given energy band. In the second panels, the slopes 
of the phase lags show no
obvious breaks; we might expect changes in the slopes if the soft and 
hard spectral components separately affected the phase lags. 
Finally, while the 
fraction of the flux in the thermal component increases in a similar manner
during the X-ray faint, radio plateau (center panel) and the X-ray faint, 
radio bright observation (left panel), the coherence function 
at low frequencies plummets dramatically with increasing energy
in the plateau observation (center panel), but is near unity when 
the X-ray flux is bright (right panel).
We conclude that the changes in the timing properties are 
not related to the fraction
of the emission which can be modeled with a thermal component. 

\subsection{Time Evolution of X-ray and Radio Properties}

\begin{figure}
\plotone{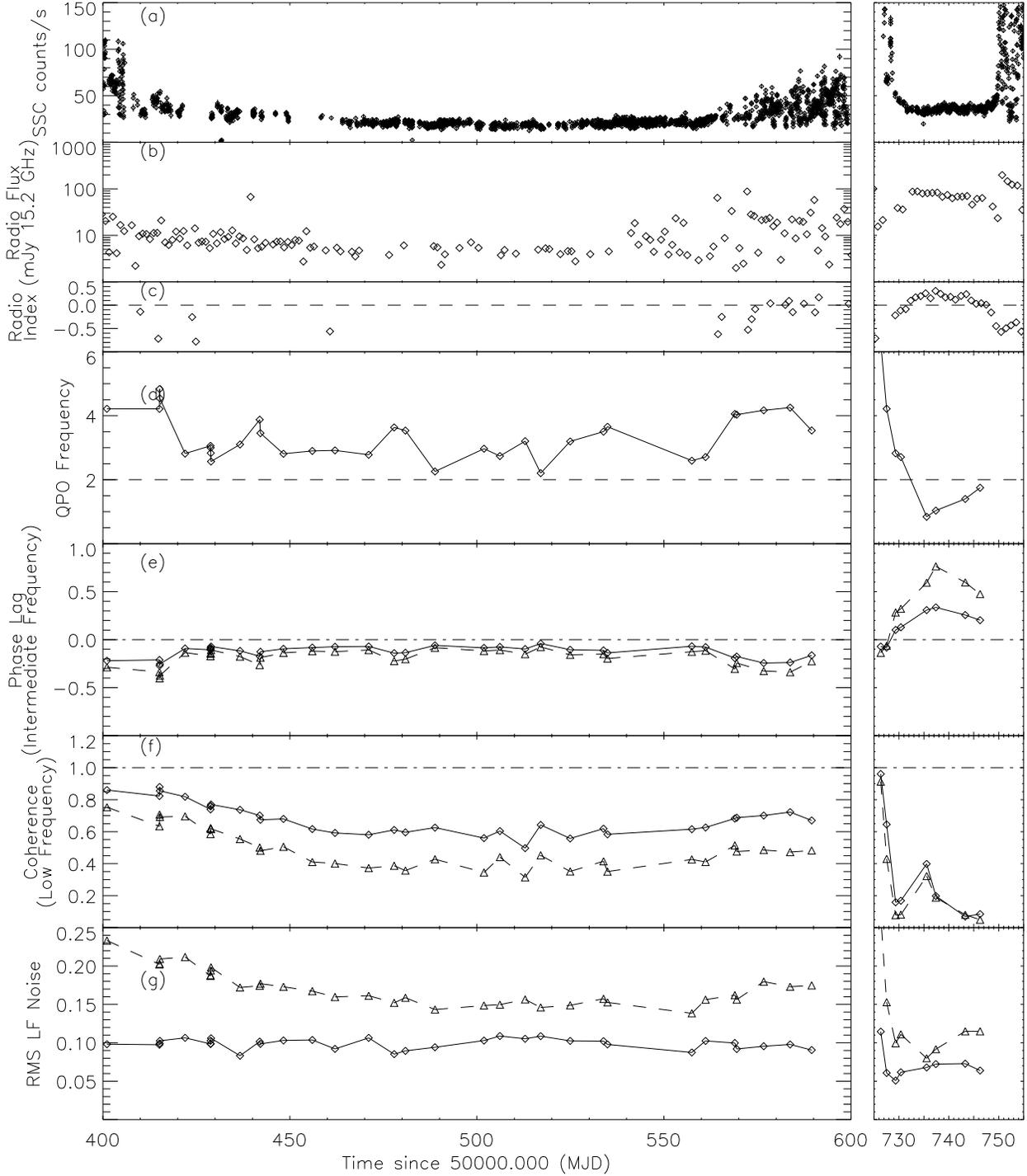}
\caption{The time evolution of the parameters of the PDS and CS for 
a stretch of faint radio emission (MJD 50400--50600, left panel) and
bright radio emission (MJD 50725--50755, right panel).
a) The ASM light curve, 1.5--12 keV. b) Ryle monitoring data (15.2 GHz).
c) The radio spectral index ($\alpha$) measured by with the GBI (see 
Figure~1).
d) The 0.5--10 Hz QPO frequency (measured from the 2--4.3~keV band). 
e) Phase lags for the 2--4.3~keV relative to the 
7.8--11.5~keV band
(diamonds connected with the solid line) and the 11.5--60~keV band
(triangles connected with the dashed line) measured between the 
low frequency break and the QPO.
f) The average coherence function for the 2--4.3~keV 
relative to the 7.8--11.5~keV band (diamonds connected with the solid 
line) and the 11.5--60~keV band (triangles connected with the dashed line)
measured below the low frequency break.
g) The RMS power at frequencies below the QPO 
measured in the 2--4.3~keV band (diamonds connected 
with the solid line) and in the 11.5--60~keV (triangles connected with
the dashed line). }
\label{lhev}
\end{figure}

\placefigure{lhev}

The evolution of the parameters of the power spectrum, phase lags, and
coherence function can be viewed as a function of time during the 
long periods in which
\grsmq\ remains in the hard-steady state.  The results demonstrate 
that the three observations selected for 
Figures~\ref{spec}--\ref{vsspec} are representative of the groups which
we have defined. In Figure \ref{lhev} we 
show two time intervals containing the hard-steady states of 
\grsmq. In the first the radio emission is faint (MJD~50400--50600, 
left panels), and the second coincides with strong radio plateau 
emission (MJD~50725--50755, right panels). 
In the top three panels from both time intervals we plot: 
a) the intensity of \grsmq\ as viewed by 
the ASM,  b) the radio flux measured with the Ryle telescope at 15.2 GHz, 
and c) the radio spectral index between 2.25 and 8.3 GHz ($\alpha$) 
measured with the GBI.
Notice that before the period of radio-plateau emission (right panels) 
the radio spectral index increases smoothly and becomes 
positive (more optically thick), and then it decreases symmetrically
afterwards. Optically thin radio flares that are 
much more prominent at 2.25 and 8.3 GHz (not shown)
occur both before and after this interval.

In panel d), we plot the frequency of the 0.5-10 Hz
QPO. \citet{tcg99} and \citet{cts97} separately established 
that the QPO frequency tracks the flux and the count rate in the PCA 
during hard-steady states similar to those displayed. 
However, it is clear that QPO frequency is 
systematically higher during radio faint emission ($> 2$ Hz) than during 
radio plateau emission ($< 2$ Hz),
despite the fact that the ASM count rate is lower during the faint radio 
emission. 

In panel e) of 
Figure~\ref{lhev} we plot the average phase lags in region~2 
between the 2--4.3~keV band and 1) the 7.8--11.5 keV band 
(diamonds connected by the
solid line), and 2) the 11.5--60 keV band (triangles connected by the
dashed line). During radio-faint hard-steady conditions, the phase lags are
consistently negative and have a value of about $-0.2$ radians. During 
radio plateau emission, the phase lags are positive and increase strongly
with energy. Only during radio plateau conditions are the X-ray phase lags
positive, as might be expected from a simple model which seeks to explain 
the timing properties using Comptonization (see Section~3).

In panel f) of Figure~\ref{lhev} we plot the average coherence 
function in region~1 between the 
2--4.3~keV band and 1) the 7.8--11.5 keV bands (diamonds connected by the
solid line), and 2) the 11.5--60 keV bands (triangles connected by the
dashed line) as a function of time.
The coherence over this frequency range drops dramatically
when the radio emission is strong and optically thick, as shown in the
representative observation of the radio plateau state in the center 
panel of Figure~\ref{pdsetc}.

In the bottom panel we plot the total of the RMS noise at frequencies
lower than the QPO (region~1 and region~2 combined). The absolute
values of the RMS low frequency noise should not be compared directly
from observation to observation, because as the QPO frequency varies, so
does the upper limit to the frequencies which we consider. However, we
have plotted the RMS low frequency noise for two energy bands: 
the 2--4.3~keV band (diamonds 
connected by a solid line) and the 11.5--60~keV band (triangles 
connected by dashed lines). It is clear that there is much more low 
frequency power in the 11.5--60~keV band than in the
2--4.3~keV band during radio-faint hard X-ray conditions, 
while during radio-plateau conditions, the amount of power in both 
energies is comparable.
This can also be seen by comparing the radio plateau and X-ray-faint,
radio-faint observations in Figure~\ref{pdsetc}.

\subsection{Timing Properties and the Radio States}

\begin{figure}
\plotone{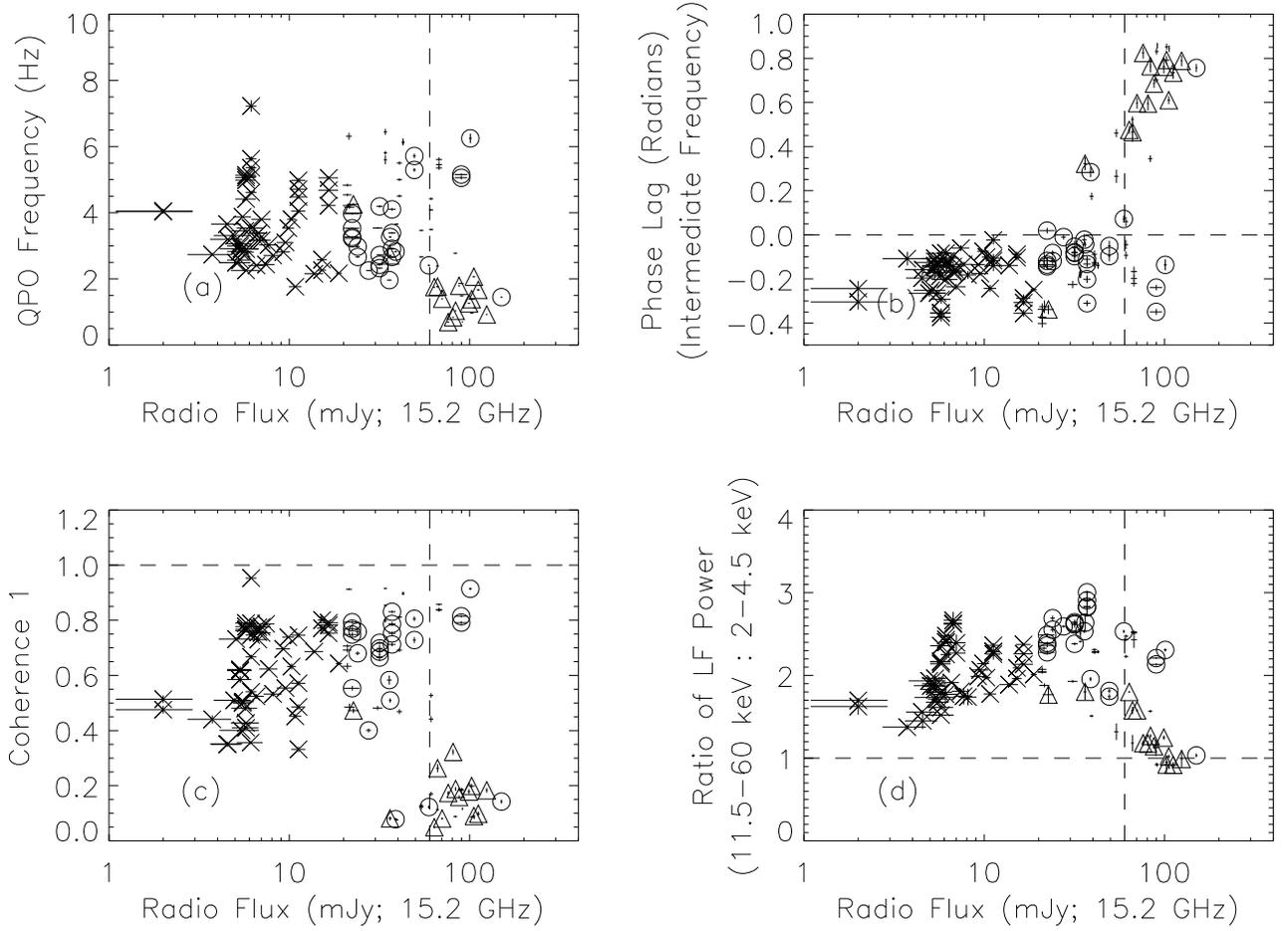}
\caption{Parameters of the PDS and CS as a function
of the radio flux (15.2 GHz). 
Triangles represent radio-plateau 
hard-steady conditions, circles radio-steep conditions, and $\times$'s
radio-faint conditions. 
a) The QPO frequency. b) Phase lags for 11.5--60~keV photons 
relative to 2--4.3~keV photons, averaged over intermediate frequencies 
(between the low frequency break and the QPO).
c) The coherence function for 2--4.3~keV photons relative to 
11.5--60~keV photons, averaged over low frequencies (below the low frequency
break). d) The ratio of the integrated broad band powers below the QPO 
frequency in the high (11.5--60 keV) to low (2--4.3 keV) energy bands.}
\label{ratfl}
\end{figure}

\begin{figure}
\plotone{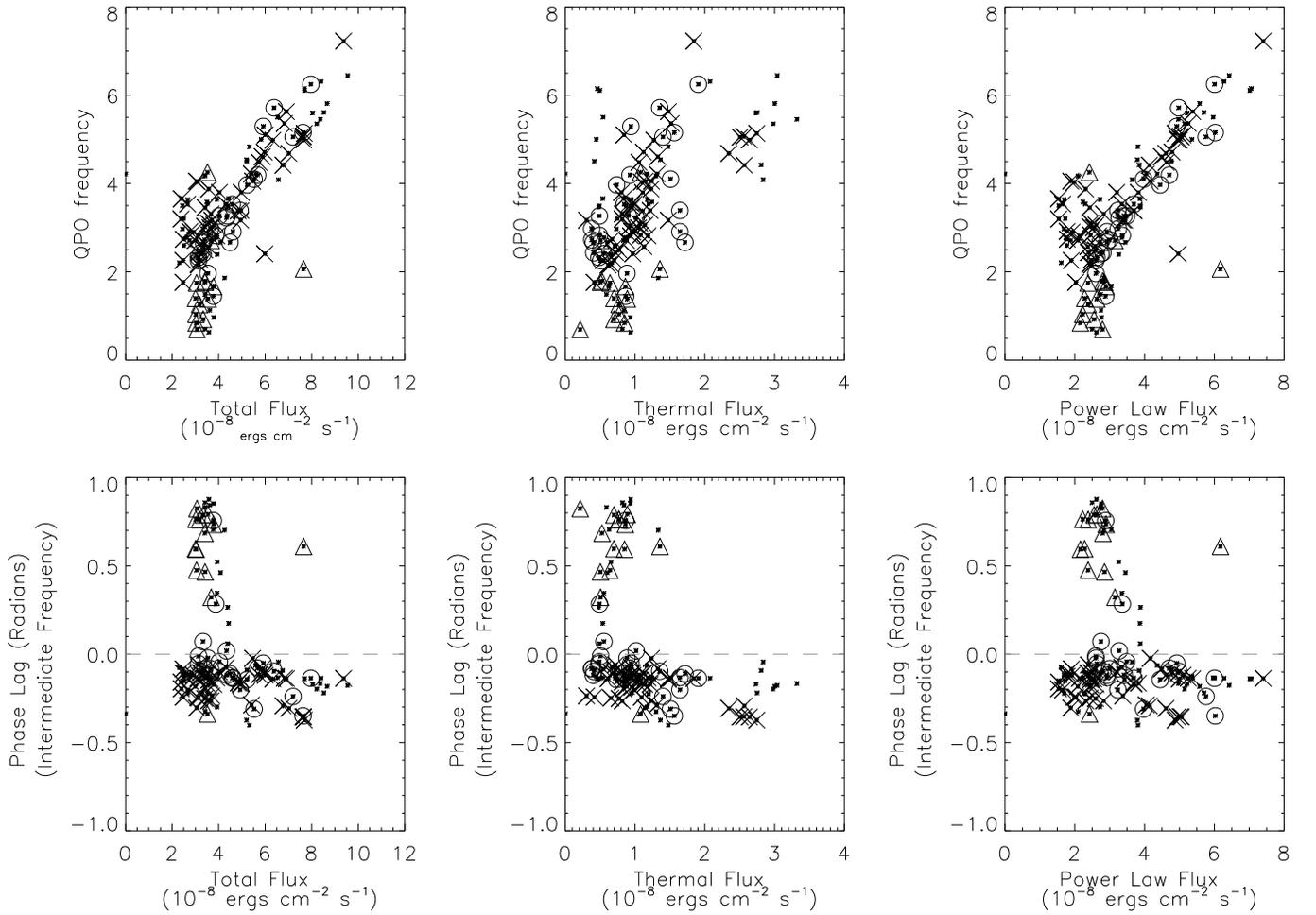}
\caption{Correlations between the QPO frequency (top panels) or the
average phase lag at intermediate frequencies (11.5--60 keV relative to
2--4.3 keV; bottom panels) with fluxes derived from spectral fits to 
the data. Left panels: total flux. Center panels: thermal flux, from
the multi-temperature disk model. Right panels: power law flux, integrated
from 2--200 keV. The symbols are the same as Figure~7.}
\label{qpocnt}
\end{figure}

We now examine directly the relationship between the radio and
the X-ray properties of \grsmq.
In Figures~\ref{ratfl} and \ref{qpocnt}, the various symbols denote the 
emission states in Figure~\ref{rvx}. 

\placefigure{ratfl}

In Figure~\ref{ratfl} we plot as a function of the 
radio flux density at 15.2 GHz: a) the QPO frequency, 
b) the phase lag between the
2--4.3~keV and 11.5--60~keV bands in region 2, 
c) the coherence for the same energy bands in region 1, and 
d) the ratio of the low-frequency (regions 1 and 2) power in the 
11.5--60~keV to 2--4.3~keV energy bands. At first glance, there appear
to be continuous distributions in the radio and timing properties in 
Figure~\ref{ratfl}, but in fact the populations of radio faint observations
and radio plateau observations are well-separated, with the radio steep
observations falling in between the two. Moreover, the $x$-axes of the
panels are plotted with a logarithmic scale, which emphasizes the large 
range in the radio flux for the various hard-steady
observations.

Four conclusions can be drawn from Figure~\ref{ratfl} regarding the changes 
in the timing properties 
as a function of the radio flux. The QPO frequency 
tends to decrease with increasing radio flux, and almost all of 
the observations 
with radio plateau conditions (triangles) have QPO frequencies less than 
2~Hz. Second, 
during radio plateau hard-steady emission the hard photons {\it lag} the 
soft, while during radio faint ($\times$'s)
conditions the hard photons 
{\it lead} the soft. Third, the coherence at low frequencies is lowest
during radio plateau emission. Finally, the ratio of low frequency power in
11.5--60 keV photons to the power in 2--4.3 keV photons is nearly 
unity only when the radio flux is largest, during
radio plateau conditions. 

\placefigure{qpocnt}

Using our large dataset which incorporates a variety of radio conditions,
 we next examine correlations which other authors have reported
to be well-defined. Two studies \citep{cts97, tcg99} 
have examined the frequency of the 0.5--10~Hz QPO as a function of
the PCA count rate (MJD~50275--50333) and the 3--20 keV X-ray flux 
(MJD~50363--50563), respectively. 
Both have found significant correlations among these values. 
In the top panels of Figure~\ref{qpocnt} we plot the QPO frequency as 
a function of total X-ray flux (defined in Section~2.3; left panel), 
thermal flux (middle panel),
and power law flux (right panel). We confirm that the frequency of the
0.5--10~Hz QPO is tightly correlated with the X-ray flux from \grsmq, 
but we also find that the correlations are not single-valued.
In the left panel there exists a significant population of points from 
radio faint observations that have QPO frequencies between 2--5~Hz, for 
which the X-ray
fluxes are low compared to those expected from the general trend. 

A plot of the average phase lags between 2--4.3 and 11.5--60 keV 
photons in region 2 as a function of the total, thermal,
and power law fluxes (bottom panels of Figure~\ref{qpocnt}) reveals 
that there is no correlation between the X-ray flux and the sign or
magnitude of the phase lags. The positive phase lags therefore appear to 
occur independently of any given X-ray flux level, and are better predicted
by the presence of optically thick radio emission. 

\subsection{Characteristics of the Faint Radio Emission}

We can use the archival VLA observations to make some conclusions 
about the nature of the faint radio emission. The multi-frequency 
radio observations of \grsmq\ are listed in Table~\ref{vla}.
The first two observations occur within a day of 
radio faint, X-ray bright observations. In both cases, the VLA observed
significant flaring, while Ryle observations, which were more nearly 
simultaneous
with the \rxte\ observations, indicate that the radio emission was 
near 5 mJy within 12 hours of the PCA observations. 
It appears that the radio emission is
unsteady within several hours of the X-ray bright hard-steady state, and we
can not rule out that the X-ray emission is also unsteady on these
time scales. Since the VLA observations show clear evidence for flaring
while the Ryle observations do not, we can not infer the radio spectrum
of the emission during the X-ray bright, hard-steady state. 

The remainder of the observations in Table~\ref{vla} are X-ray faint. There
is clear evidence in the GBI data for $\simeq$30 mJy optically thin 
flares on 1996 December 
28, 1997 January 12, and 1998 October 08. The flares decay on time scales
of several hours, and no \rxte\ observations were taken within
0.5 day of these flares. Since GBI flux measurements below 20 mJy are
uncertain, there remain two observations during which the VLA observations
demonstrate that the radio emission was faint and steady. On 1998 September 
14, 15.2 GHz Ryle and 5.0 GHz VLA observations separated by 6 hours provide
a spectral index $\alpha = -0.5\pm0.1$, which suggests that the emission 
is optically thin. However, it is certainly possible that the radio flux varied
by a few mJy during those several hours. Dual frequency VLA observations 
were made at 5.0 and 15.2 GHz on 1998 September 29, which indicate 
$\alpha = 0.12\pm0.01$.
This suggests that the emission is optically thick, similar to a weak
version of the radio plateau state. However, with only a single secure 
measurement of $\alpha$ during the radio faint state, more observations 
are certainly warranted.

\section{Discussion}

We begin the discussion with a summary of our three hard-steady conditions:

\begin{itemize}
\item {\bf Radio Faint} (e.g. MJD~50450--50550; MJD~50708) The radio
emission has a mean flux density of about 5 mJy at 15.2 GHz.
When the X-ray flux is faint ($< 4 \times 10^{-8}$ \ergcms), 
the X-ray spectrum is best described as the sum of a cool 
($T_{\rm app}<$2~keV, $R_{\rm app} \sim 60$~km)
multi-temperature blackbody and a power law of index $\Gamma \simeq 2.1$ with
a cut-off near 70 keV. 
When the X-ray emission is brighter ($> 4 \times 10^{-8}$ \ergcms), the
X-ray spectrum is best described with a hot ($>$2~keV) multi-temperature 
black body and a steep power law ($\Gamma \simeq 3.2$) without a cut off
below 200 keV (see Section~2.2).
The QPO frequency is greater than 2~Hz, phase lags are negative (or, soft,
indicating that hard photons precede soft photons)
at intermediate continuum frequencies, 
the coherence at low frequencies in the continuum is $\simeq$ 0.5--1.0 
(11.5--60 keV compared to 2--4.3 keV), and there is much more low frequency
power in the high energy band than the low energy band.

\item {\bf Radio Plateau} (e.g. MJD~50730--50748) The radio emission is bright 
($\sim$100~mJy at 15~GHz) and optically thick ($\alpha > -0.2$). 
The X-ray spectrum is best described by a hot multi-temperature black body
($T_{\rm app} > 3$~keV and $R_{\rm app} < 5$~km) and a power law 
($\Gamma \simeq 2.5$) which could be cut-off at $E > 80$~keV.
The QPO frequency reaches its lowest values ($<2$ Hz), phase lags
are positive (hard photons lag soft photons), the coherence is 
low (particularly at low frequencies),
and there is about equal broad band RMS power at low frequencies in 
all energy bands.

\item {\bf Radio Steep} (e.g MJD~50727-50730) These observations 
represent the transition into and out of radio plateau conditions. The
radio flux is bright, while the radio spectrum 
is optically thin ($\alpha < -0.2$). The X-ray flux is often higher
than during radio plateau conditions, but the spectrum is still best
described by a hot multi-temperature black body ($T_{\rm app} > 3$~keV 
and $R_{\rm app} < 10$~km) and a steep power law ($\Gamma \simeq 2.5$)
with a cut-off above 80~keV.
The timing properties tend to be similar to radio faint conditions
(although there are exceptions): the QPO frequency is
between 2--5~Hz, the phase lags are negative, the coherence is around 
0.5, and there is more broad band power at low frequencies in the
highest energy band. 

\end{itemize}

Previous observations have established the nature of the bright radio
emission from \grsmq. The optically thick emission
during plateau conditions has been resolved as a compact jet
of relativistic electrons \citep{dmr00}, similar to ones which may be present
in \cygxone\ \citep{sti98} and \gxbh\ \citep{cor00}. Our observations
of the faint radio emission provide some evidence that it is similar to 
a weak radio plateau state.
Radio flares with steep spectra originate from material which has been ejected 
from the central source \citep{fen99b, dmr00}, and so is most 
likely decoupled from the instantaneous conditions of the X-ray emitting 
regions. 
In the following discussion we will focus on comparing the
radio faint and radio plateau conditions, during which the radio and X-ray
properties of \grsmq\ are closely related.

The timing properties of \grsmq--- the QPO frequency, phase lags,
coherence function, and broad band noise--- appear to be highly dependent 
upon the radio conditions (Figure~\ref{ratfl}). However,
it is not immediately obvious whether the changes in the timing properties are
part of the mechanism which produces a compact jet, or 
whether they result from changes in the accretion flow induced when 
the compact radio jet is strong. Since the location and origin 
of the Comptonizing electrons which generate hard X-rays is unknown, 
the possible connection between these and the jet is doubly
uncertain. Nevertheless, we may explore interpretations of these results 
using simple models for the accretion flow under consideration
in the literature.

In Section~4.1, we consider the extent to which Comptonization of a
soft input signal by a static corona at the base of a compact radio 
jet is consistent with the phase lags and X-ray spectrum of \grsmq.
In Sections~4.2--4.3, we examine two generic models which incorporate more
elements of the accretion flow. In both models, cool 
($T \sim 1$ keV) thermal emission originates from an optically thick 
accretion disk, while the hot power law component originates from
inverse Compton scattering. The first basic 
model places the Comptonizing electrons in a spherical corona within 
the inner radius of the accretion disk, while the second geometry 
assumes that the relativistic electrons are part of  
a planar corona sustained by magnetic flares above the disk. 

\subsection{Generic Comptonization}

Hard phase lags are generic properties of the low-hard X-ray state of 
black hole candidates such as \cygxone\ \citep{now99}
and \gxbh\ \citep{nwd99}. Surprisingly, 
the phase lags in \grsmq\ change from negative to positive as the
strength of the radio emission increases (Figure~\ref{ratfl}b).
Since \citet{fen99a} have suggested that the 
Comptonizing corona in the hard state of black hole candidates 
represents the base of a compact radio jet, we feel it is important
to explore the possibility that the negative phase lags during radio faint 
observations of \grsmq\ are intrinsic to the region close to the black 
hole, while
the hard phase lags generated during the radio plateau state are due to  
the Comptonization of the intrinsic input signal in a large corona at 
the base of the compact radio jet.

The hard phase lags in \grsmq\ may constrain the structure of 
the Comptonizing region, because the phase is roughly constant as a 
function of frequency (Figure~\ref{pdsetc}). This is contrary
to the constant time lags that would be expected from a homogeneous corona.
Instead, \citet{kht97} and 
B\"ottcher \& Liang (1999) have demonstrated that constant phase lags can 
be produced in an isotropically illuminated, spherical Compton cloud with 
a uniform temperature and a density which decreases with radius as $R^{-1}$,
since photons have an equal probability of scattering per decade of radius
in such a cloud (see Nowak et al. 1999b 
for further discussion of these models). 
High-frequency variability originates from photons which 
have scattered from small radii, since the signal is washed out if 
it scatters with long time delays. On the other hand, 
the time lags of low-frequency signals are dominated by photons which 
have scattered over large radii. This produces time lags which decrease
as a function of frequency, or phase lags which are nearly constant
as a function of frequency.

Several authors have pointed out that the phase lags which are observed 
during the hard states of black hole candidates imply 
large scales for inverse Compton scattering, which raises the question of 
how energy is transported to support a large, hot corona 
\citep{pfa99, now99b}.
In the plateau state of \grsmq\, a phase lag of 0.5 radians at 0.3~Hz 
(middle panel of Figure~\ref{pdsetc}) would imply a light travel time of 
0.3 s, or a distance of $R\tau \simeq 8\times10^9$ cm (where $\tau$ is the
optical depth of the corona). For a 10 \msun\ black 
hole, this represents thousands of Schwarschild radii. However, this 
distance is much smaller than the size of the compact jet observed
by Dhawan et al. (2000), 10 AU $= 5\times10^{14}$ cm, which is consistent 
with the assumption that the Comptonizing electrons reside at the base of 
the jet. If the Compton cloud is the base of a compact jet, this 
reduces the difficulty in explaining how energy is transported 
to support a very large corona, 
as the jet may carry up to 10$^{39}$ erg s$^{-1}$ of power to much 
larger radii \citep{fb99}.

It is possible that the radio faint states are associated with a smaller
Compton corona located near the inner accretion disk. In this case, 
negative phase lags could be caused by any number of effects intrinsic
to the accretion disk, such 
as waves similar to those of \citet{now99} which 
propagate outward in the disk rather than inward, or a modified form the of 
the magnetic flares of \citet{pfa99}.
Even if a scattering corona intercepts a significant fraction of the photons
from the disk, negative phase lags could be
observed so long as 1) the scattering time in the corona 
is much smaller than the intrinsic time lags \citep{mil95} and
2) the differences between the energies of the seed photons and the 
observed signals are not too large (which should be the case in \grsmq, since
the apparent disk temperature is $> 0.8$~keV; Nowak \& Vaughan 1996).

Moreover, the radius of the Compton corona can change without greatly affecting
the slope and cut-off energy of the power law. 
The electron temperature ($kT_e$)
of the corona is constrained to be nearly constant by the cut-off in 
the power law at 70--80 keV measured during X-ray faint observations
(left and center panels of Figure~\ref{spec}). If the temperature
of the corona is fixed, a steeper power law  
(higher $\Gamma$) implies a lower optical depth \citep[see][]{st80}.
Since the power law is steeper during radio plateau observations
($\Gamma \simeq 2.5$) than during the radio faint, X-ray faint 
observations ($\Gamma \simeq 2.1$), we can suggest that the optical depth 
of the corona decreases and its radius increases as the compact radio 
jet becomes more luminous. However, the X-ray bright, radio faint observations
would then have a small, optically thin corona, since the phase
lags are negative and the power law is steep ($\Gamma \simeq 3.2$) during
these observations (Figures~\ref{spec} and \ref{pdsetc}). The electron 
acceleration mechanism could be different in the X-ray bright hard 
state, as is hinted at by the absence of an observable cutoff in the 
power law below 200~keV.

As others have noted, estimates of the Lorentz factors 
for electrons emitting synchrotron emission 
\citep[$\sim 400$;][]{fb99} suggest energies that are much larger than that
of the Comptonizing electrons. One may consider the 
acceleration mechanism for the jet as an undetermined structure
that draws material from the Compton corona near the inner disk, as
in \citet{fen99a}. In the following two sections, we
therefore briefly speculate on the implications which the simplest 
geometries for the Compton corona described in the literature have on 
the jet production mechanism.

\subsection{A Spherical Corona within the Inner Accretion Disk}

Many authors have suggested that a 
spherical corona is formed within the inner radius of an
accretion disk, either as an advection-dominated accretion flow (ADAF; 
Esin, McClintock, \& Narayan 1997), or by a shock 
in the accretion flow \citep{ct95}. 
The compact jet responsible for the radio emission
could form from a wind emanating from an ADAF \citep{bb99}
or a post-shock region \citep{cha99}. 
The distinctive feature of the spherical corona model is that the
radius of the inner accretion disk is variable, which
leads naturally to a scenario in which the X-ray timing and spectral 
properties, and the strength of the radio emission, are set by the inner 
radius of the accretion disk. 

The X-ray timing property that is perhaps most relevant to discuss 
in the spherical corona model is the 0.5--10~Hz QPO. The QPO may represent
a characteristic time scale in the accretion 
flow such as an acoustic \citep[e.g.][]{ct94, act95}, Keplerian 
\citep{rod00}, 
or free-fall (Molteni, Sponholz, \& Chakrabarti 1996) 
time scale. Since the QPO peak is narrow ($Q > 3$), it must originate 
in a localized region, as all of the time scales increase as the 
radius increases. Among the possibilities for the origin of such a QPO are
oscillations in a shock in the accretion flow \citep{cm00} or a spiral 
density wave in the inner disk \citep{rod00}. 
Since the frequency of the QPO decreases during episodes in which the 
inner radius of the disk is thought to increase \citep{bel97b, mst99}, 
it seems plausible that the frequency of the 0.5--10 Hz 
QPO in \grsmq\ tracks the inner radius of the accretion disk. 
Figure~\ref{ratfl}a then implies 
that the radio emission is strongest when the inner disk is farthest 
from the source. 

Nobili et al. (2000) have developed a model for 
the phase lags in \grsmq\ by assuming a spherical corona with 
two temperature regions: a warm ($T \simeq 1.5$ keV) outer corona and 
a hot ($T \simeq 15$ keV) inner corona. The corona is isotropically illuminated
by soft seed photons. The sign and magnitude of the phase
lags are correlated with the radius of the inner accretion disk, 
because as the disk moves inward it compresses the corona, increasing its
optical depth. Large, positive phase lags occur when the inner disk
radius is large, because the hot corona has an optical depth 
$\tau_{\rm H} \sim 1$, 
while the warm corona is optically thin. When the inner radius is a
factor of $\sim 3$ smaller, the optical depth of the 
hot corona becomes extremely high ($\tau_{\rm H} > 100$), and the soft seed 
photons from the disk thermalize to the temperature of the optically 
thick hot corona. The thermalized photons are then Compton 
down-scattered in the cool corona, which produces soft lags. 
This model does not attempt to explain the phase lags as a function of
Fourier frequency, but it does predict that if the QPO 
frequency tracks the radius of the inner disk, negative phase lags are expected
when the QPO frequency is high--- exactly as Reig et al (2000) observed in 
\grsmq. However, a corona with an optical depth of $\tau > 100$ would 
produce a thermal spectrum, while 
the energy spectrum of \grsmq\ between 25--200~keV can not be 
described by an extremely optically thick spectrum (e.g. a spherical 
Compton cloud 
with optical depth $\tau > 5$; Sunyaev \& Titarchuk 1980) 
when the phase lags are negative. 
Further work needs to be done to reconcile the negative phase lags with
the spectral shape under this type of model.

The interpretation of the coherence function and the continuum power
as a function of energy is less certain. When the QPO frequency 
is low, the coherence at low frequencies is nearest to 0 (Figure~\ref{ratfl}c)
and the ratio of the low frequency continuum power at high to 
low energies is near 1 (Figure~\ref{ratfl}d).
A coherence less than 1 can occur if the variability is the 
superposition of separate linear signals with different properties 
\citep{vn97}. When
the inner disk is farthest from the source (low QPO frequency) there
may be more area available for multiple inputs to contribute to the
variability. 

The radius of the inner disk should also determine the energy spectrum
under a spherical corona model. We have already discussed the model
of \citet{nob00}. A different set of assumptions are used by,
for example \citet{emn97} and \citet{ct95},
who take into account the thermal balance between the soft photons and the
Comptonizing electrons. Soft spectra are formed when the 
inner radius of the disk is close to the last stable orbit, so that the
disk produces copious amounts of seed photons which cool the corona. Hard
states are formed when the inner radius of disk is far from the last stable
orbit and most of the corona is photon-starved. However, if we 
consider the faint X-ray observations, the power law is steeper
($\Gamma \simeq 2.5$) when the QPO frequency is lower and the disk 
is inferred to be farther from the black hole (radio plateau observations; 
Figure~\ref{spec} and \ref{pdsetc}). 
This is opposite the sense expected, because the Compton corona should
be cooled less efficiently (which implies a lower $\Gamma$) when the 
inner disk is large.  If the QPO frequency tracks the 
transition radius between the disk and the corona, there must an additional 
factor beside the radius of the inner accretion disk which affects the 
observed X-ray properties of the corona in \grsmq\ under this model.

\subsection{A Planar Corona Sustained by Magnetic Flares}

In a planar corona model, one can assume that magnetic flares 
powered by the differential rotation of the accretion disk produce a 
population of relativistic electrons above the disk (di Matteo et al.
2000; compare Dove et al. 1997 for difficulties with
static models). 
The compact jet could represent a collimated wind which forms according to the 
magnetic-centrifugal mechanism of \citet{bp82}, and 
is strongest when the flares provide electrons enough energy to 
escape the potential of the black hole \citep{rom98}.
In the magnetic flare model of \citet{dim99}, 
the inner radius of the accretion disk in general does not vary.  

A natural starting point for discussing the planar corona models are the
phase lags observed in \grsmq. The time lags which would be implied by the 
scale height of the flares in the model of \citet{dim99} are too 
short ($10^7$ cm $= 10^{-3}$ s) to explain the phase lags observed in 
black hole candidates. However, \citet{pfa99} have developed 
a model for \cygxone\ in which the spectral evolution of the flares which 
accelerate Comptonizing electrons produce the observed power density
spectra and phase lags. 
Their model produces only hard (or zero) lags, because the 
electrons are assumed to be in thermal equilibrium during the flare, so
the spectrum of the flares is set by the ratio of 
the heating in the flares to the amount of soft seed flux impinging on the
flares. The 
flares become harder because the majority of the soft flux is produced
by feedback in response to the flare, and the feedback is taken 
to decrease as a function of time as the flares move away from the disk. 
However, if the electrons accelerated in the magnetic flare have a 
non-thermal energy distribution which evolves as a function of time,
it is conceivable that either hard or soft phase lags could be 
produced.
The hard phase lags may represent the most efficient acceleration
of the electrons, which could in turn explain why the radio plateau
emission is associated with the largest hard 
phase lags. On the other hand, the negative phase lags could represent
spectral evolution in which the soft energy band peaks later, and relativistic 
electrons are accelerated less efficiently. 

In the case of a magnetically sustained planar corona with a 
fixed inner disk radius, 
there is some difficulty in explaining what sets the frequency of the 
QPO and why this frequency varies. Since the power in the X-ray variability 
in \grsmq\ drops sharply at frequencies higher than the 0.5--10~Hz QPO
(Figure~\ref{pdsetc}), 
we can speculate this QPO represents a resonance at the minimum time scale at 
which magnetic flares can form. The radio emission, then, would occur 
most favorably when only relatively slow flares form (low QPO frequency).

The coherence function and the ratio of the high to low energy broad-band
power would also be set by the evolution of the magnetic flares. Since
the coherence at low frequencies is lowest when the QPO frequency is 
low and the phase lags are positive, the long, hard flares may 
undergo significant non-linear spectral evolution. 
The ratio of broadband power in the high
and low energy bands could be set by the shape of individual flares in the 
respective energy bands. Mathematically, this could be described in a manner
similar to the shot noise models of \citet{shi88}.

The model of \citet{dim99} predicts the energy spectrum from 
magnetic flares, by assuming that a fixed fraction of the accretion power 
is released into the magnetic structures, and that the electrons which are
accelerated by the flares are cooled inverse-Compton scattering of 
both cyclo-synchrotron and thermal (disk) seed photons. 
Soft spectral states are caused by flares 
which occur close to the disk and are flooded by seed photons, while 
hard states are caused by photon-starved flares which occur high above 
the disk. Although \citet{dim99} do not attempt to explain the 
phase lags observed in black hole candidates, we might expect hard phase
lags to correspond with flares that produce the hardest (lowest $\Gamma$) 
power law, since the flares would be highest above the disk and cooled 
least efficiently. However, when
the phase lags are positive in \grsmq\ we find a 
steeper power law ($\Gamma \simeq 2.5$ in the plateau state) than when 
the phase lags are 
negative ($\Gamma \simeq 2.1$ during radio faint, X-ray faint observations). 
Clearly, simple assumptions about the evolution of magnetic flares would
have to be 
relaxed in order to explain the X-ray timing and spectral properties
of \grsmq.

\section{Conclusions}

We have found that although radio emission is always associated with the
hard-steady state of \grsmq\ (between 1996-1999), 
the 15.2 GHz radio flux density and the
2--200~keV X-ray flux are not correlated. Instead, we find that the 
hard states can be described using three basic regimes of radio emission:
a) radio plateau conditions in which bright, 
optically thick emission originates 
from a compact jet, b) radio steep conditions in which bright, optically thin 
emission originates from material most likely decoupled from the X-ray
emitting region, and c) faint radio conditions for which our current study 
is unable to constrain the spectrum of the radio emission. As 
the radio flux increases, 1) the frequency of a ubiquitous 
0.5--10 Hz QPO decreases, 2) the Fourier phase lags between hard 
(11.5--60 keV) and soft (2--4.3 keV) photons in the frequency range of 
0.01--10 Hz change sign from negative to positive, 3) the coherence at 
low frequencies decreases, and 4) the relative amount of low 
frequency power in hard photons compared to soft 
photons decreases. 

We have attempted to understand these results qualitatively using 
simple models collected from the literature, but we find that our
attempts to understand the X-ray timing properties of the hard state 
of \grsmq\ inevitably lead to contradictions with the observed energy 
spectrum. These observational results therefore provide impetus for 
developing more sophisticated models of the accretion flow and the 
compact jet in \grsmq.

\acknowledgments

Radio astronomy at the Naval
Research Laboratory is supported by the Office of Naval Research.
The Green Bank
Interferometer is a facility of the National Science Foundation,
operated by the National Radio Astronomy Observatory in support of the
NASA High Energy Astrophysics programs.
The NRAO VLA is a facility of the National Science Foundation,
operated under Cooperative Agreement by Associated Universities, Inc.
This work was also supported in part by NASA contract NAS 5-30612.
Robert Hjellming passed away in July 2000, and his contributions to 
astrophysics will be sorely missed.

\begin{deluxetable}{llccccc}
\tabletypesize{\small}
\tablecolumns{7}
\tablewidth{0pc}
\tablecaption{Hard-Steady Pointed {\it RXTE} Observations of 
GRS~1915$+$105\label{rxte}}
\tablehead{
\colhead{ObsID} & \colhead{Date} & \colhead{X-ray}
& \colhead{HR\tablenotemark{b}} &\colhead{$S_{2.25}$} & \colhead{$S_{8.3}$} 
& \colhead{$S_{15.2}$} \\
\colhead{} & \colhead{(UT)} & \colhead{Flux\tablenotemark{a}}
& \colhead{} & \colhead{(mJy)} & \colhead{(mJy)} 
& \colhead{(mJy)}
}
\startdata
10408-01-22-00 & 1996 Jul 11 02:09 & 4.15(1) & 0.063 & \nodata & \nodata &  54(2) \\
10408-01-22-01 & 1996 Jul 11 05:16 & 3.93(1) & 0.070 & \nodata & \nodata &  54(2) \\
10408-01-22-02 & 1996 Jul 11 08:38 & 3.78(1) & 0.073 & \nodata & \nodata &  54(2) \\
10408-01-23-00 & 1996 Jul 14 11:45 & 4.17(1) & 0.063 & \nodata & \nodata &  39.3(6) \\
10408-01-24-00 & 1996 Jul 16 04:04 & 3.62(1) & 0.075 & \nodata & \nodata &  66(1) \\
10408-01-25-00 & 1996 Jul 19 11:45 & 3.20(1) & 0.101 & \nodata & \nodata & 105.9(7) \\
10258-01-01-00\tablenotemark{c} & 1996 Jul 23 10:33 & 3.54(1) & 0.118 & \nodata & \nodata & 65.8(7) \\
10408-01-27-00 & 1996 Jul 26 13:55 & 3.12(1) & 0.114 & \nodata & \nodata &   \nodata \\
10258-01-02-00 & 1996 Jul 29 18:28 & 3.00(1) & 0.111 & \nodata & \nodata &  91.1(6) \\
10408-01-28-00 & 1996 Aug 03 12:43 & 3.32(1) & 0.105 & \nodata & \nodata & 103(2) \\
10258-01-03-00 & 1996 Aug 06 07:12 & 3.12(1) & 0.096 & \nodata & \nodata &  90.3(4) \\
10408-01-29-00 & 1996 Aug 10 08:52 & 3.59(1) & 0.090 & \nodata & \nodata &  89.4(6) \\
10258-01-04-00 & 1996 Aug 14 12:14 & 3.66(1) & 0.075 & \nodata & \nodata &  83.3(5) \\
10408-01-30-00 & 1996 Aug 18 07:26 & 5.01(1) & 0.058 & \nodata & \nodata &  40.8(6) \\
10258-01-05-00 & 1996 Aug 20 07:40 & 7.46(1) & 0.055 & \nodata & \nodata &  42.7(5) \\
10408-01-31-00 & 1996 Aug 25 04:47 & 5.15(1) & 0.065 & \nodata & \nodata &  61.1(5) \\
10258-01-06-00 & 1996 Aug 29 11:31 & 8.66(1) & 0.058 & \nodata & \nodata & 35.5(8) \\
10408-01-32-00 & 1996 Aug 31 07:55 & 8.03(1) & 0.058 & \nodata & \nodata &  34.1(4) \\
10408-01-33-00 & 1996 Sep 07 18:00 & 6.71(1) & 0.063 & \nodata & \nodata &  67.6(5) \\
10408-01-42-00 & 1996 Oct 23 03:36 & 6.37(1) & 0.079 & \nodata & \nodata &  16.6(4) \\
10408-01-43-00 & 1996 Oct 23 12:00 & 5.85(1) & 0.079 & \nodata & \nodata &  16.6(4) \\
10408-01-45-00 & 1996 Oct 29 12:00 & 4.21(1) & 0.096 & \nodata & \nodata &   6.7(6) \\
20402-01-01-00 & 1996 Nov 07 05:45 & 5.98(1) & 0.078 & \nodata & \nodata & 31.8(8) \\
20402-01-02-01 & 1996 Nov 14 02:09 & 4.76(1) & 0.090 & \nodata & \nodata &  16.5(5) \\
20402-01-02-02 & 1996 Nov 14 15:07 & 5.76(1) & 0.077 & \nodata & \nodata & 4.3(9) \\
20402-01-04-00 & 1996 Nov 28 03:21 & 4.95(1)  & 0.086 & \nodata & \nodata & 21(1) \\
20402-01-05-00 & 1996 Dec 04 23:31 & 3.33(1) & 0.116 & \nodata & \nodata &   9.2(6) \\
20402-01-06-00 & 1996 Dec 11 18:43 & 3.23(1) & 0.119 & 10(4) & 7(6) &   5.3(7) \\
20402-01-07-00 & 1996 Dec 19 15:50 & 3.04(1) & 0.115 & 24(4) & 11(6) &   9.5(6) \\
20402-01-08-00 & 1996 Dec 24 22:04 & 2.96(1) & 0.102 & 16(4) & 10(6) &   5.5(5) \\
20402-01-08-01 & 1996 Dec 25 02:52 & 2.94(1) & 0.113 & 15(4) & 8(6) &   5.7(7) \\
20402-01-09-00 & 1996 Dec 31 06:43 & 2.64(1) & 0.124 & 16(4) & 5(6) &   5.5(7) \\
20402-01-10-00 & 1997 Jan 07 23:45 & 2.44(1) & 0.125 & 5(4) & 5(6) &   5.8(8) \\
20402-01-11-00 & 1997 Jan 14 01:26 & 2.29(1) & 0.124 & 12(4) & 7(6) &   5(1) \\
20402-01-12-00 & 1997 Jan 23 01:40 & 2.20(1) & 0.125 & 9(4) & 7(6) &   \nodata \\
20402-01-13-00 & 1997 Jan 29 20:52 & 2.20(1) & 0.110 & 7(4) & 7(6) &   \nodata \\
20402-01-14-00 & 1997 Feb 01 21:07 & 2.18(1) & 0.111 & 10(4) & 5(6) &   6.1(8) \\
20402-01-15-00 & 1997 Feb 09 18:43 & 2.20(1) & 0.144 & \nodata & \nodata &   5.8(7) \\
20402-01-16-00 & 1997 Feb 22 21:07 & 2.08(1) & 0.131 & 9(4) & 7(6) &   \nodata \\
20402-01-17-00 & 1997 Feb 27 03:36 & 2.15(1) & 0.131 & 13(4) & 7(6) &   3.7(8) \\
20402-01-18-00 & 1997 Mar 05 21:21 & 2.13(1) & 0.120 & 10(4) & 8(6) &   \nodata \\
20402-01-19-00 & 1997 Mar 10 01:12 & 1.98(1) & 0.146 & 13(4) & 6(6) &   \nodata \\
20402-01-20-00 & 1997 Mar 17 22:04 & 1.98(1) & 0.121 & 9(4) & 5(6) &   4.6(8) \\
20402-01-21-00 & 1997 Mar 26 20:09 & 2.01(1) & 0.114 & 10(4) & 9(6) &   \nodata \\
20402-01-21-01 & 1997 Mar 27 21:36 & 1.98(1) & 0.110 & 16(4) & 6(6) &   4.5(7) \\
20402-01-25-00 & 1997 Apr 19 10:47 & 2.16(1) & 0.133 & 6(4) & 7(6) &   \nodata \\
20402-01-24-00 & 1997 Apr 23 03:07 & 2.33(1) & 0.132 & 9(4) & 7(6) &   \nodata \\
20402-01-26-01 & 1997 Apr 30 22:33 & 2.48(1) & 0.100 & 8(4) & 10(6) &   2.0(7) \\
20402-01-26-02 & 1997 May 01 10:47 & 2.55(1) & 0.103 & 8(4) & 9(6) &   2.0(7) \\
20402-01-27-01 & 1997 May 08 16:04 & 2.77(1) & 0.097 & 15(4) & 14(6) &  22(1) \\
20402-01-27-03\tablenotemark{c} & 1997 May 15 05:16 & 2.43(1) & 0.113 & 22(4) & 46(6) & \nodata \\
20187-02-02-00 & 1997 May 15 11:31 & 2.98(1) & 0.097 & 27(4) & 30(6) &  22.7(8) \\
20402-01-29-00 & 1997 May 21 11:45 & 2.93(1) & 0.107 & 13(4) & 11(6) &  30.9(8) \\
20186-03-02-05 & 1997 Sep 17 14:09 & 8.45(1) & 0.057 & 11(4) & 11(6) &   6.1(3) \\
20186-03-02-06 & 1997 Sep 18 03:07 & 5.70(1) & 0.062 & 10(4) & 10(6) &  11.2(3) \\
20402-01-47-01\tablenotemark{c}& 1997 Sep 19 00:00 & 6.14(1) & 0.055 & 15(4) & 14(6) &  19.3(6) \\
20187-02-03-00 & 1997 Oct 03 20:38 & 7.82(1) & 0.063 & 280(7)& 108(7)&  101(1) \\
20187-02-04-00 & 1997 Oct 05 08:23 & 7.37(1) & 0.057 & 32(4) & 13(6) &  21.4(6) \\
20187-02-05-00 & 1997 Oct 06 11:45 & 5.01(1) & 0.062 & 32(4) & 16(6) &   \nodata \\
20187-02-06-00 & 1997 Oct 07 10:05 & 5.84(1) & 0.060 & 25(4) & 12(6) & \nodata \\
20402-01-49-00 & 1997 Oct 08 07:55 & 3.63(1) & 0.080 & 33(4) & 22(6) &  38.8(5) \\
20402-01-49-01 & 1997 Oct 09 09:21 & 3.43(1) & 0.082 & 44(4) & 35(6) &  36.2(5) \\
20402-01-50-00 & 1997 Oct 14 13:12 & 2.61(1) & 0.124 & 50(4) & 65(6) &  80.8(5) \\
20402-01-50-01 & 1997 Oct 16 09:36 & 2.64(1) & 0.117 & 50(4) & 74(6) &  83.5(6) \\
20402-01-51-00 & 1997 Oct 22 06:57 & 2.66(1) & 0.113 & 45(4) & 65(6) &  70.3(7) \\
20402-01-52-00 & 1997 Oct 25 06:28 & 2.73(1) & 0.108 & 46(4) & 41(6) &  63.6(5) \\
30703-01-14-00 & 1998 Apr 06 21:07 & 3.35(1) & 0.092 & 78(4) & 91(6) & 111.6(9) \\
30402-01-09-00 & 1998 Apr 09 21:07 & 3.54(1) & 0.087 & 87(4) & 85(6) & 105(1) \\
30402-01-09-01 & 1998 Apr 10 00:00 & 3.58(1) & 0.086 & 87(4) & 85(6) & 105(1) \\
30402-01-10-00 & 1998 Apr 11 09:21 & 3.46(1) & 0.089 & 100(4) & 122(7) &   \nodata \\
30402-01-11-00 & 1998 Apr 20 06:14 & 5.45(1) & 0.052 & 188(5) & 80(6) &  49.3(6) \\
30703-01-15-00 & 1998 Apr 22 21:21 & 3.34(1) & 0.094 & 132(4) & 100(6) & 150(1) \\
30703-01-16-00 & 1998 Apr 28 16:19 & 3.10(1) & 0.101 & 82(4) & 111(7) & 102.5(6) \\
30703-01-17-00 & 1998 May 05 21:36 & 2.99(1) & 0.112 & 145(5) & 116(7) & 124(1) \\
30402-01-12-00 & 1998 May 11 22:33 & 3.18(1) & 0.093 & 70(4) & 87(6) &  87.3(4) \\
30402-01-12-01 & 1998 May 12 00:28 & 3.15(1) & 0.098 & 70(4) & 87(6) &  87.3(4) \\
30402-01-12-02 & 1998 May 12 02:09 & 3.17(1) & 0.099 & 70(4) & 87(6) &  87.3(4) \\
30402-01-12-03 & 1998 May 12 03:50 & 3.17(1) & 0.101 & 70(4) & 87(6) &  87.3(4) \\
30703-01-19-00 & 1998 May 19 14:23 & 2.79(1) & 0.112 & 99(4) & 95(6) &  99(1) \\
30703-01-20-00 & 1998 May 24 19:26 & 2.96(1) & 0.135 & 49(4) & 53(6) &  76.0(9) \\
30703-01-21-00 & 1998 May 31 19:26 & 3.17(1) & 0.098 & 52(4) & 46(6) &  66.3(8) \\
30402-01-14-00 & 1998 Jun 11 08:09 & 3.05(1) & 0.096 & 104(4) & 53(6) &  59.4(6) \\
30703-01-18-00 & 1998 Jun 16 18:14 & 3.79(1) & 0.087 & 91(4) & 43(6) &  36.2(5) \\
30703-01-22-00 & 1998 Jun 27 14:38 & 2.93(1) & 0.105 & 46(4) & 27(6) &  27.5(5) \\
30703-01-22-01 & 1998 Jun 27 16:33 & 2.94(1) & 0.106 & 46(4) & 27(6) &  27.5(5) \\
30182-01-01-01 & 1998 Jul 08 04:04 & 2.79(1) & 0.125 & 87(4) & 33(6) &  35.8(9) \\
30182-01-01-00 & 1998 Jul 08 05:02 & 2.80(1) & 0.116 & 87(4) & 33(6) &  35.8(9) \\
30182-01-02-01\tablenotemark{c} & 1998 Jul 9 04:04 & 3.92(1) & 0.109 & 200(6) & 97(6) &  22.3(9) \\
30182-01-02-00 & 1998 Jul 09 05:02 & 4.16(1) & 0.098 & 200(6) & 97(6) &  22.3(9) \\
30182-01-03-01\tablenotemark{c} & 1998 Jul 10 04:05 & 7.84(1) & 0.074 & 430(7)  & 137(9) & 152(1) \\
30182-01-03-00 & 1998 Jul 10 10:05 & 7.10(1) & 0.079 & 350(7) & 139(9) & 90(1) \\
30182-01-04-00 & 1998 Jul 11 05:45 & 4.77(1) & 0.090 & 143(5) & 47(6) &  37.1(4) \\
30182-01-04-02 & 1998 Jul 12 04:04 & 3.46(1) & 0.102 & 93(4) & 38(6) &  23.9(5) \\
30182-01-04-01 & 1998 Jul 12 05:02 & 3.15(1) & 0.103 & 93(4) & 38(6) &  23.9(5) \\
30703-01-25-00 & 1998 Jul 23 18:00 & 3.56(1) & 0.103 & 28(4) & 12(6) &   7.0(7) \\
30402-01-16-00 & 1998 Aug 28 16:47 & 3.02(1) & 0.118 & 16(4) & 11(6) &   5.1(9) \\
30703-01-31-00 & 1998 Aug 31 18:28 & 3.63(1) & 0.106 & 40(4) & 28(6) &  10.3(5) \\
30402-01-17-00 & 1998 Sep 11 15:07 & 3.35(1) & 0.112 & 17(4) & 10(6) &   \nodata \\
30703-01-33-00 & 1998 Sep 15 22:47 & 3.27(1) & 0.113 & \nodata & \nodata &   5(1) \\
30703-01-34-00 & 1998 Sep 22 00:14 & 2.27(1) & 0.146 & 13(4) & 15(6) &  10.8(7) \\
30703-01-35-00 & 1998 Sep 25 19:40 & 2.79(1) & 0.136 & 12(4) & 9(6) &   7.4(7) \\
30703-01-36-00 & 1998 Oct 03 17:02 & 2.94(1) & 0.125 & 28(4) & 10(6) &   7.7(7) \\
30402-01-18-00 & 1998 Oct 10 08:23 & 2.56(1) & 0.114 & 15(4) & 9(6) &   8.1(8) \\
30703-01-24-01 & 1998 Dec 18 16:47 & 2.92(1) & 0.129 & 12(4) & 13(6) &   6.6(6) \\
30703-01-24-02 & 1998 Dec 18 19:26 & 2.77(1) & 0.129 & 12(4) & 13(6) &   6.6(6) \\
30703-01-24-00 & 1998 Dec 19 03:21 & 2.81(1) & 0.135 & 12(4) & 13(6) &   6.6(6) \\
30703-01-41-00 & 1998 Dec 26 16:04 & 2.83(1) & 0.139 & 16(4) & 9(6) &  14(1) \\
40703-01-01-00 & 1999 Jan 01 01:26 & 2.86(1) & 0.136 & 37(4) & 16(6) &  15.2(7) \\
40703-01-02-00 & 1999 Jan 08 01:40 & 3.85(1) & 0.111 & \nodata & \nodata & 9.9(6) \\
40703-01-03-00 & 1999 Jan 15 23:45 & 2.80(1) & 0.139 & 21(4) & 30(6) &  18.8(6) \\
40403-01-01-00 & 1999 Jan 24 22:04 & 2.98(1) & 0.129 & 17(4) & 14(6) &  15.2(8) \\
40703-01-04-02 & 1999 Feb 06 08:09 & 2.88(1) & 0.100 & 12(4) & 14(6) &   \nodata \\
40703-01-05-00 & 1999 Feb 12 01:12 & 3.71(1) & 0.101 & 68(4) & 65(6) & 47.3(8) \\
\enddata
\tablenotetext{a}{Between 2--200 keV, in units of 
$10^{-8}$ erg cm$^{-2}$ s$^{-1}$.}
\tablenotetext{b}{Ratio of PCA counts from 12--60 : 2--12 keV.}
\tablenotetext{c}{Observations which were not analyzed either because they 
were not taken with the necessary data modes or were too short for meaningful
timing analyses.}
\end{deluxetable}

\begin{deluxetable}{llcccc}
\tabletypesize{\small}
\tablecolumns{6}
\tablewidth{0pc}
\tablecaption{Radio Observations of GRS 1915+105 while Faint \label{vla}}
\tablehead{
\colhead{Date} & \colhead{Time} & \colhead{Instrument} & 
\colhead{Frequency} & \colhead{$S_\nu$} & \colhead{Beam}\\
\colhead{} & \colhead{(UT)} & \colhead{} &
\colhead{GHz} & \colhead{mJy} & \colhead{(arcsec)}
}
\startdata
1996 Oct 29 & 01:15 -- 02:56 & VLA & 8.4  & 30 -- 70 & 0.26 $\times$ 0.23 \\ 
1996 Oct 29 & 16:37 -- 19:56 & Ryle & 15.5 & 6.7 & 30 \\ [10pt]

1996 Nov 14 & 23:37 -- 01:21 & VLA & 8.4 & 18 -- 8 & 0.27 $\times$ 0.24 \\ 
1996 Nov 14 & 13:34 -- 18:38 & Ryle & 15.5 & 4.3 & 30 \\ [10pt]

1996 Dec 28 & 20:50 -- 23:00 & VLA & 8.4 & 4(1) & 0.32 $\times$ 0.26 \\
1996 Dec 28 & 20:50 -- 23:00 & GBI & 2.25 & 17(4) & 11 \\ 
1996 Dec 28 & 20:50 -- 23:00 & GBI & 8.3 & 14(6) & 3 \\ 
1996 Dec 28 & 09:30 -- 12:30 & Ryle & 15.2 & 6.4(4) & 30 \\[10pt]

1997 Jan 12 & 21:50 -- 22:30 & VLA & 4.8 & 35 -- 26 & 0.63 $\times$ 0.36 \\
1997 Jan 12 & 21:50 -- 22:30 & VLA & 8.4 & 30 -- 19  & 0.41 $\times$ 0.29 \\
1997 Jan 12 & 12:45 -- 20:57 & GBI & 2.25 & 44(4) & 11 \\ 
1997 Jan 12 & 12:45 -- 20:57 & GBI & 8.3 & 21(6) & 3 \\[10pt]

1997 Jan 13 & 16:50 -- 18:35 & VLA & 8.4 & 4.5(1) & 0.27 $\times$ 0.24 \\
1997 Jan 13 & 16:50 -- 18:35 & GBI & 2.25 & 14(4) & 11 \\
1997 Jan 13 & 16:50 -- 18:35 & GBI & 8.3 & 4(6) & 3 \\[10pt]

1997 Apr 19 & 10:40 -- 15:20 & VLA & 1.4 & 3.9(1) & 4.5 $\times$ 3.5\\
1997 Apr 19 & 10:40 -- 15:20 & GBI & 2.25 & 6(4) & 11 \\
1997 Apr 19 & 10:40 -- 15:20 & GBI & 8.3 & 6(6) & 3 \\[10pt]

1998 Sep 14 & 04:20 -- 05:30 & VLA & 8.4 & 8.2(1) & 0.74 $\times$ 0.70\\
1998 Sep 13 & 23:07 -- 04:26 & GBI & 2.25 & 18(4) & 11 \\
1998 Sep 13 & 23:07 -- 04:26 & GBI & 8.3 & 11(6) & 3 \\
1998 Sep 13 & 21:39 -- 22:39 & Ryle & 15.2 & 6.0(6) & 30 \\[10pt]

1998 Sep 29 & 04:00 -- 06:30 & VLA & 5.0 & 12.2(1) & 2.26 $\times$ 1.2 \\ 
1998 Sep 29 & 04:00 -- 06:30 & VLA & 15.2 & 13.9(1) & 0.40 $\times$ 0.38 \\ 
1998 Sep 28 & 21:36 -- 03:36 & GBI & 2.25 & 20(4) & 11 \\ 
1998 Sep 28 & 21:36 -- 03:36 & GBI & 8.3 & 12(6) & 3 \\
1998 Sep 28 & 19:07 -- 21:30 & Ryle & 15.2 & 10.7(4) & 30 \\[10pt]

1998 Oct 08 & 03:30 -- 05:30 & VLA & 5.0 & 31.7(1) &  1.6 $\times$ 1.3 \\
1998 Oct 08 & 03:30 -- 05:30 & VLA & 15.2 & 18.5(1) & 0.42 $\times$ 0.41 \\ 
1998 Oct 07 & 21:22 -- 03:07 & GBI & 2.25 & 60(4) & 11 \\
1998 Oct 07 & 21:22 -- 03:07 & GBI & 8.3 & 23(6) & 3 \\
1998 Oct 07 & 19:05 -- 21:19 & Ryle & 15.2 & 21.3(4) & 30 \\
\enddata
\end{deluxetable}

\end{document}